\documentclass[a4paper,UKenglish,cleveref,pdfa]{lipics-v2021}
\nolinenumbers
\ArticleNo{1}
\hideLIPIcs
\usepackage{amssymb,amsmath,amsthm,mathtools}
\usepackage{tikz}
\usetikzlibrary{calc,positioning}
\usepackage{tikz-cd}
\usepackage[inline]{enumitem}
\newcommand{\TA}{{\mathsf{TA}}}

\newcommand{\Sig}{\mathsf{Sig}}
\newcommand{\Mod}{\mathsf{Mod}}
\newcommand{\Sen}{\mathsf{Sen}}
\newcommand{\Class}{\mathbb{C}\mathsf{lass}}
\newcommand{\Set}{\mathbb{S}\mathsf{et}}
\newcommand{\A}{\mathfrak{A}}
\newcommand{\B}{\mathfrak{B}}

\newcommand{\D}{\mathfrak{D}}

\newcommand{\M}{\mathcal{M}}

\newcommand{\K}{\mathcal{K}}

\renewcommand{\S}{\mathcal{S}}
\renewcommand{\P}{\mathcal{P}}
\newcommand{\frag}{\mathcal{L}}
\newcommand{\card}{\mathsf{card}}
\newcommand{\act}{\mathfrak{a}}
\newcommand{\Space}{~~~~~}

\newcommand{\suc}{\mathrel{\vartriangleleft}}

\newcommand{\pos}[1]{{\langle}#1{\rangle}}

\newcommand{\Forall}[1]{\forall #1\,{\cdot}\,}
\newcommand{\Exists}[1]{\exists #1\,{\cdot}\,}
\newcommand{\Frac}[2]{\displaystyle\frac{#1}{#2}}
\newcommand{\red}{\mathord{\upharpoonright}}
\newcommand{\sep}{~\mid~}
\usepackage{scalerel}
\newcommand{\bbsemicolon}{%
  \scalerel*{%
    \hbox{\usefont{U}{bbold}{m}{n} ;}%
  }{;}%
}
\newcommand{\comp}{\mathbin{\bbsemicolon}}
\usepackage{longtable}
\usepackage{array,etoolbox}
\makeatletter

\newlength{\PS@lastparam}
\newlength{\PSlastparam}
\newcommand{\PSlp}{%
  \setlength{\PSlastparam}{\PS@lastparam}%
  \the\PSlastparam
}
\def\PS@sub@lastparam{}

\newcommand{\PS@numwidth}{99}
\newcommand{\PSnumwidth}[1]{%
  \renewcommand{\PS@numwidth}{#1}%
}

\newcommand{\PS@style}{\small}
\newcommand{\PS@numstyle}{\footnotesize}

\newlength{\PSindent}
\newlength{\PS@extraindent}
\newlength{\PSpre}
\newlength{\PSpost}
\newlength{\PS@Nwidth}
\newlength{\PS@Swidth}
\newlength{\PS@Ewidth}
\newlength{\PScolsep}
\newcommand{\PS@rownumber}{%
  \ifPS@subsubsteps
  \thePSsubstepc.%
  \the\numexpr\value{PSsubsubstepc}+1\relax
  \else
  \ifPS@substeps
  \thePSstepc.%
  \the\numexpr\value{PSsubstepc}+1\relax
  \else
  \the\numexpr\value{PSstepc}+1\relax
  \fi\fi
}
\newcommand{\PS@step}{%
  \ifPS@subsubsteps
  \refstepcounter{PSsubsubstepc}%
  \else
  \ifPS@substeps
  \refstepcounter{PSsubstepc}%
  \else
  \refstepcounter{PSstepc}
  \fi\fi%
}

\newif\ifPS@inprogress
\newif\ifPS@substeps
\newif\ifPS@subsubsteps
\newif\ifPS@continued
\newif\ifPS@subcontinued
\newcounter{PSc}
\newcounter{PSstepc}[PSc]
\newcounter{PSsubstepc}[PSstepc]
\renewcommand{\thePSsubstepc}{\thePSstepc.\arabic{PSsubstepc}}
\newcounter{PSsubsubstepc}[PSsubstepc]

\newenvironment{proofsteps}[1]{%
  \global\settowidth{\PS@lastparam}{\PS@style\hspace*{#1}}
  \ifPS@continued\else\refstepcounter{PSc}\fi
  \begingroup
  \setlength{\LTpre}{\PSpre}%
  \setlength{\LTpost}{\PSpost}%
  
  \setlength{\tabcolsep}{0pt}
  \noindent\PS@style
  \settowidth{\PS@Nwidth}{\PS@numstyle\PS@numwidth}%
  \setlength{\PS@Swidth}{#1}%
  \addtolength{\PS@Swidth}{-\PS@extraindent}%
  \setlength{\PS@Ewidth}{\linewidth}%
  \addtolength{\PS@Ewidth}{-\PSindent}%
  \addtolength{\PS@Ewidth}{-\PS@extraindent}%
  \addtolength{\PS@Ewidth}{-\PS@Nwidth}%
  \addtolength{\PS@Ewidth}{-\PScolsep}%
  \addtolength{\PS@Ewidth}{-\PS@Swidth}%
  \addtolength{\PS@Ewidth}{-\PScolsep}%
  \PS@inprogresstrue
  \longtable{%
    @{\hspace*{\PSindent}\hspace*{\PS@extraindent}\makebox[\PS@Nwidth][r]{\PS@rownumber}}%
    @{\hskip\PScolsep}>{\PS@step}p{\PS@Swidth}%
    @{\hskip\PScolsep}>{\footnotesize\raggedright\arraybackslash}p{\PS@Ewidth}%
  }%
}{%
  \ifPS@inprogress
  \addtocounter{table}{-1}%
  \endlongtable  
  \endgroup
  \PS@continuedfalse
  \PS@inprogressfalse
  \else\fi
}

\newcommand{\PSbreak}[1]{%
  \endproofsteps
  \par\medskip
  #1
  \medskip\par
  \PS@continuedtrue
  \proofsteps{\PS@lastparam}%
}

\newif\ifPS@sub@inprogress

\newif\ifPS@laststep
\newcommand{\laststep}{\global\PS@laststeptrue}
\newif\ifPS@lastsubstep
\newcommand{\lastsubstep}{\global\PS@lastsubsteptrue}

\newcommand{\adjustcol}[1]{%
  \global\advance\@colroom-#1%
}

\makeatother
\title{Model-theoretic Forcing in Transition Algebra}
\author{Go Hashimoto}%
{Kyushu University, Fukuoka, Japan}%
{go.427@s.kyushu-u.ac.jp}%
{https://orcid.org/0009-0000-2722-0204}%
{}

\author{Daniel G\u{a}in\u{a}}%
{Kyushu University, Fukuoka, Japan}%
{daniel@imi.kyushu-u.ac.jp}%
{https://orcid.org/0000-0002-0978-2200}%
{}

\authorrunning{G.~Hashimoto and D.~G\u{a}in\u{a}}
\Copyright{G.~Hashimoto and D.~G\u{a}in\u{a}}
\funding{The work presented in this paper has been partially supported by Japan Society for the Promotion of Science, grant number 23K11048.}

\keywords{institutional model theory, algebraic specification, transition algebra, forcing, omitting types property, L\" owenheim-Skolem properties, completeness}
\EventEditors{Paweł Gawrychowski, Filip Mazowiecki, and Michał Skrzypczak}
\EventNoEds{3}
\EventLongTitle{50th International Symposium on Mathematical Foundations of Computer Science (MFCS 2025)}
\EventShortTitle{MFCS 2025}
\EventAcronym{MFCS}
\EventYear{2025}
\EventDate{August 25--29, 2025}
\EventLocation{Warsaw, Poland}
\EventLogo{}
\SeriesVolume{345}
\ArticleNo{87}
\begin{document}

\maketitle

\begin{abstract}
We study L\" owenheim-Skolem and Omitting Types theorems in Transition Algebra, a logical system obtained by enhancing many sorted first-order logic with features from dynamic logic. 
The sentences we consider include compositions, unions, and transitive closures of transition relations, which are treated similarly to actions in dynamic logics to define necessity and possibility operators.
We show that Upward L\" owenheim-Skolem theorem, any form of compactness, and joint Robinson consistency property fail due to the expressivity of transitive closures of transitions. 
In this non-compact many-sorted logical system, we develop a forcing technique method by generalizing the classical method of forcing used by Keisler to prove Omitting Types theorem. 
Instead of working within a single signature, we work with a directed diagram of signatures,  which allows us to establish Downward  L\" owenheim-Skolem and Omitting Types theorems despite the fact that models interpret sorts as sets, possibly empty.
Building on a complete system of proof rules for Transition Algebra, we extend it with additional proof rules to reason about constructor-based and/or
finite transition algebras. We then establish the completeness of this extended system for a fragment of Transition Algebra obtained by restricting models to constructor-based and/or finite
transition algebras.
\end{abstract}

%%% %%% %%% %%% %%% %%% %%% %%% %%%
\section{Introduction}

Transition Algebra~\cite{go-icalp24}, abbreviated $\TA$, offers a logical framework for modeling and reasoning about state transitions in systems, particularly in concurrency theory and computer science. Closely related to modal logics and algebraic specification languages executable by rewriting, it provides a structured approach to describing system evolution based on predefined rules.
The key concepts in transition algebra are 
\begin{enumerate*}[label=(\alph*)]
\item states, which are characterized using equational theories, and 
\item transitions, which are represented as labeled relations between states.
\end{enumerate*}
Transition algebra can be broadly viewed as an extension of many-sorted first-order logic, incorporating features from dynamic logic. It includes algebraic operators that define how transitions combine, such as
\begin{enumerate*}[label=(\alph*)]
\item Sequential composition ($\comp$): 
if a system moves from state $t_1$ to $t_2$ according to transition rule $\act_1$, and then from  $t_1$ to $t_2$ under transition rule $\act_2$, this is written as $\act_1\comp\act_2$,
\item Choice ($\cup$): represents nondeterministic choices between transitions. 
\item Iteration ($*$):  denotes repeated transitions, analogous to the Kleene star in formal language theory.
\end{enumerate*}
Iteration ($*$), has been employed in the literature to enhance the expressive power of first-order logic, albeit at the expense of compactness.
Notably, Fagin’s work on Transitive Closure Logic (TCL) has been extensively studied within the frameworks of finite model theory and descriptive complexity~\cite{fagin1974spectra,Fagin1993}.
However, apart from the work on $\TA$~\cite{go-icalp24}, we are not aware of any model-theoretic studies of first-order logic with iteration beyond the scope of finite model theory. Moreover, when sorts—possibly infinitely many—and models interpreting those sorts as sets—possibly empty—are permitted, extending existing techniques to such a framework poses significant challenges and lacks a straightforward solution.

In \cite{go-icalp24}, it is shown that transition algebra possesses an essential property for modularization, known as the \emph{satisfaction condition}. This condition states that truth remains invariant under changes in notation. Here, a change in notation refers to the translation between local languages constructed from vocabularies, often called signatures. Consequently, transition algebra is formalized as an institution \cite{gog-ins}, a category-theoretic framework that captures the notion of logic by incorporating its syntax, semantics, and the satisfaction relation between them. 
Similarly, the logic underlying CafeOBJ~\cite{dia-caf} satisfies the satisfaction condition for signature morphisms, allowing it to be formalized as an institution. However, its models cannot distinguish between different transitions occurring between the same states, as transitions are represented by preorder relations. In contrast, Rewriting Logic~\cite{Meseguer90} -- the foundation of Maude~\cite{conf/maude/2007} -- offers greater expressivity by treating transitions as arrows in a category, where states serve as objects. This expressiveness comes at a cost, as the satisfaction condition does not hold in Rewriting Logic. On the other hand, transition algebra retains the modular properties of the preorder algebras in CafeOBJ due to its institutional structure while achieving the same level of expressivity as Rewriting Logic.

Given a set of properties formalized as first-order sentences, practitioners seek to study a restricted class of transition algebras that satisfy these properties. In fields such as formal methods and functional programming—where computation and reasoning rely on abstract data types and term rewriting—many algebraic structures are naturally described using constructor operators, such as numbers with 0 and succ, or lists with nil and cons. Pattern matching on constructors further simplifies function definitions.
A related example is Finite Model Theory~\cite{Libkin2004,Gradel2007-GRDFMT}, which focuses on logical structures with a finite domain. This area is particularly relevant to database theory, complexity theory, and various computer science applications.
Building on a complete system of proof rules for $\TA$, such as the one proposed in \cite{go-icalp24}, we extend it with additional proof rules to reason about constructor-based and/or finite transition algebras. 
We refer the reader to \cite{DBLP:journals/tcs/BidoitHK03,DBLP:journals/jlp/BidoitH06,gai-cbl} for theoretical foundations of constructor-based models and to \cite{riescoACM25} for a formal method developed for reasoning about such models.
We then establish the completeness of this extended system for the fragment of $\TA$ obtained by restricting the semantics to constructor-based and/or finite transition algebras.

In this work, we use Omitting Types Theorem (OTT) to extend completeness of $\TA$ to a fragment obtained by restricting the semantics to constructor-based transition algebras with finite carrier sets.
The OTT states that if a first-order theory $\Phi$ does not require a certain type (a set of formulas describing a potential element) to be realized,
then there exists a model of $\Phi$ that omits this type.
This ensures the existence of models that exclude specific unwanted elements.
A key challenge arises from the fact that models with potentially empty carrier sets are allowed in $\TA$.
In this model-theoretic framework, 
extending the initial signature with an infinite number of constants cannot always be achieved while preserving the satisfiability of the underlying theory. 
To address this, we employ a forcing technique introduced in \cite{go-icalp24} to prove completeness of $\TA$.
Forcing was originally introduced by Paul Cohen~\cite{cohen63,cohen64} in set theory to show the independence of the continuum hypothesis from the other axioms of Zermelo-Fraenkel set theory. 
Robinson~\cite{rob71} developed an analogous theory of forcing in model theory.
The forcing technique has been studied in institutional model theory at an abstract, category-theoretic level, in works such as~\cite{gai-comp,gai-ott,gai-acm}.
The forcing method proposed in \cite{go-icalp24} generalizes classical forcing from a single signature to a category of signatures,
enabling the dynamic addition of both constants and sentences while preserving satisfiability.
Notably, this dynamic forcing method is particularly suited for logical systems that are not compact, such as $\TA$.

Additionally, we show that $\TA$ theories have some control over the cardinality of their models. The Upward L\"owenheim–Skolem Theorem does not hold in general, as there exist theories that uniquely determine a model up to isomorphism, provided that the model's size is exactly the cardinality of the power of its underlying signature.\footnote{The power of a signature is the cardinality of the set of all sentences that can be formed in its language.} 
It is interesting to note that the Downward L\"owenheim–Skolem Theorem does hold, as for any model whose size exceeds the power of its underlying signature, one can construct an elementary submodel whose size matches the power of the signature.
%%% %%% %%% %%% %%% %%% %%% %%% %%%
\section{Transition algebra} \label{section:transition-algebra}
%%% %%% %%% %%% %%% %%% %%% %%% %%%
In this section, we recall \emph{the logic of many-sorted transition algebras}~\cite{go-icalp24}, or \emph{transition algebra}~($\TA$), for short.
We present, in order: signatures, models, sentences, and the \(\TA\) satisfaction relation.
%%% %%% %%%
\subparagraph*{Signatures}
%%% %%% %%%
The signatures we consider are ordinary algebraic signatures endowed with polymorphic transition labels.
We denote them by triples of the form \((S,F, L)\), where:
\begin{itemize}[topsep=3pt]

\item \((S, F)\) is a many-sorted algebraic signature consisting of a set of \emph{sorts} \(S\)  and a set of \emph{function symbols} \(F = \{ \sigma:w\to s \mid w \in S^{*}\text{ and } s \in S \}\); 

\item \(L\) is a set whose elements we call \emph{transition labels}.
\end{itemize}
Given a function symbol \(\sigma:w\to s\in F\), we refer to \(w \in S^*\) as its \emph{arity} and \(s \in S\) as its \emph{sort}.
When \(w\) is the empty arity, we may speak of \(\sigma \colon \to s\) as a \emph{constant} of sort \(s\).

Throughout the paper, we let \(\Sigma\), \(\Sigma'\), and \(\Sigma_{i}\) range over signatures of the form \((S, F, L)\), \((S', F', L')\), and \((S_{i},F_{i},L_{i})\), respectively.

As usual in institution theory~\cite{Diaconescu2008,SannellaTarlecki2011}, important constructions such as signature extensions with constants as well as open formulae and quantifiers are realized in a multi-signature setting, so moving between signatures is common.
A \emph{signature morphism} \(\chi \colon \Sigma \to \Sigma'\) consists of a triple $\chi=(\chi^{st},\chi^{op},\chi^{lb})$, where
\begin{enumerate*}[label=(\alph*)]
\item $\chi^{st}:S\to S'$ is a function mapping sorts, 
\item $\chi^{op}:F\to F'$ maps each function symbol $\sigma:s_1\dots s_n \to s\in F$ to a function symbol $\chi^{op}(\sigma):\chi^{st}(s_1)\dots\chi^{st}(s_n)\to \chi(s)\in F'$ and 
\item $\chi^{lb}:L\to L'$ is a function mapping labels.
\end{enumerate*}
For convenience, we typically omit the superscripts $st$, $op$ and $lb$ in the notation.

\begin{remark}
Signature morphisms compose componentwise. 
This composition is associative and has identities, forming a category $\Sig$ of signatures.
\end{remark}
%%% %% %%%
\subparagraph*{Models}
%%% %% %%%
Given a signature \(\Sigma\), a \emph{\(\Sigma\)-model} \(\mathfrak{A}\) consists of: 
\begin{itemize}[topsep=3pt]
\item an \((S, F)\)-algebra \(\mathfrak{A}\), where each sort $s\in S$ is interpreted by $\A$ as a set $\A_s$ and each function symbol $\sigma:s_1\dots s_n\to s\in F$ is interpreted by $\A$ as a function $\sigma^\A:\A_{s_1}\times\ldots\times \A_{s_n}\to \A_s$;
\item an interpretation of each label \(\varrho \in L\) as a \emph{many-sorted transition relation} $\varrho^\A \subseteq \A \times \A$, where  $\varrho^\A=\{\varrho^\A_s\}_{s\in S}$ and $\varrho^\A_s\subseteq \A_s\times \A_s$ for all sorts $s\in S$.
\footnote{In principle, relations of any arity could be defined, and the results presented in this paper would still hold. However, in algebraic specification languages that are executable via rewriting, the only relations considered are transitions, which are defined between pairs of elements of the same sort. Furthermore, the iteration operator ($*$) can be applied exclusively to transitions.}
\end{itemize}
%%%
A \emph{homomorphism} $h:\A\to \B$ over a signature $\Sigma$ is an algebraic \((S, F)\)-homomorphism that preserves transitions: $h(\varrho^\A)\subseteq \varrho^\B$ for all \(\varrho \in L\).
It is easy to see that $\Sigma$-homomorphisms form a category, which we denote by $\Mod^\TA(\Sigma)$, under their obvious componentwise composition.
%%%
\begin{remark}
Every signature morphism \(\chi \colon \Sigma \to \Sigma'\) determines a \emph{model-reduct functor} $\_\red_\chi\colon\Mod^\TA(\Sigma')\to\Mod^\TA(\Sigma)$ such that:
\begin{itemize}[topsep=3pt]
\item for every \(\Sigma'\)-model \(\A'\), 
\((\A'\red_\chi)_s=\A'_{\chi(s)}\) for each sort \(s \in S\), 
\(\sigma^{(\A'\red_{\chi})} = \chi(\sigma)^{\A'}\) for each symbol \(\sigma \in F\), and 
\(\varrho^{(\A'\red_{\chi})} = \chi(\varrho)^{\A'}\) for each label \(\varrho \in L\); and

\item for every \(\Sigma'\)-homomorphism \(h' \colon \A' \to \B'\), \((h' \red_{\chi})_{s} = h'_{\chi(s)}\) for each \(s \in S\).
\end{itemize}
Moreover, the mapping \(\chi \mapsto \_\red_\chi\) is functorial.
\end{remark}
%%%
For any signature morphism $\chi:\Sigma\to\Sigma'$, any $\Sigma$-model $\A$ and any $\Sigma'$-model $\A'$ if $\A=\A'\red_\chi$, we say that $\A$ is the \emph{$\chi$-reduct} of $\A'$, and that $\A'$ is a \emph{$\chi$-expansion} of  $\A$.
For example, for a many-sorted set \(X\) (say, of variables) that is disjoint from the set of constants in \(\Sigma\), consider the inclusion morphism $\iota_X:\Sigma\hookrightarrow\Sigma[X]$, where \(\Sigma[X] = (S, F[X],L)\) is the signature obtained from \(\Sigma\) by adding the elements of \(X\) to \(F\) as new constants of appropriate sort.
Then an expansion of a $\Sigma$-model $\A$ along $\iota_X$ can be seen as a pair $\A'=\pos{\A,f:X\to \A}$, where $f$ is a valuation of $X$ in $\A$.

As in many-sorted algebra, there is a special, initial model in \(\Mod^\TA(\Sigma)\), which we denote by \(T_{\Sigma}\), whose elements are ground terms built from function symbols, and whose transitions are all empty.
The \(\Sigma\)-model \(T_{\Sigma}(X)\) of terms with variables from \(X\) is defined as the $\iota_X$-reduct of $T_{\Sigma[X]}$; i.e., $T_\Sigma(X)=T_{\Sigma[X]}\red_{\iota_X}$.
%%% %%% %%%
\subparagraph*{Sentences}
%%% %%% %%%
Given a signature $\Sigma$,
the set $A$ of \emph{actions} is defined by the following grammar:
\begin{center}
$\act ~\Coloneqq~
\varrho \sep
\act \comp \act \sep
\act \cup \act \sep
\act^*$
\end{center}
where $\varrho$ is a transition label.
We let $A$ denote the set of all actions and extend our notational convention for signature components to their corresponding sets of actions. Specifically, we use:
\begin{enumerate*}[label=(\alph*)]
\item \(A'\)  to denote the set of actions over a signature \(\Sigma'\),  
\item \(A_{i}\)  to denote the set of actions over a signature \(\Sigma_{i}\),
\end{enumerate*}
and similarly for other variations.
Moreover, through a slight abuse of notation, we also denote by \(\chi \colon A \to A'\) the canonical map determined by a signature morphism \(\chi \colon \Sigma \to \Sigma'\).

To define sentences, we assume a countably infinite set of \emph{variable names} $\{v_i\mid i<\omega\}$.
A \emph{variable} for a signature $\Sigma$ is a triple $\pos{v_i,s,\Sigma}$, where 
\begin{enumerate*}[label=(\alph*)]
\item $v_i$ is a variable name and 
\item $s$ is a sort in $\Sigma$.
\end{enumerate*} 
The third component is used only to ensure that variables are distinct from the constants declared in \(\Sigma\), which is essential when dealing with quantifiers.
The set $\Sen^\TA(\Sigma)$ of \emph{sentences} over $\Sigma$ is given by the following grammar:
\begin{center}
  $\phi \Coloneqq  t = t' \mid \act(t_1,t_2)\mid \neg\phi \mid \lor\Phi \mid \Exists{X}\phi'$
\end{center}
where
\begin{enumerate*}[label=(\alph*)]
\item $t$ and $t'$ are $\Sigma$-terms of the same sort;
\item $\act$ is any action;
\item $t_1$ and $t_2$  are $\Sigma$-terms of the same sort;
\item $\Phi$ is a finite set of $\Sigma$-sentences; and 
\item $X$ is a finite block of $\Sigma$-variables, that is, $X$ is a finite set of variables such that if $\pos{v_i,s_1,\Sigma},\pos{v_j,s_2,\Sigma}\in X$ and $s_1\neq s_2$ then $i\neq j$; and
\item $\phi'$ is a $\Sigma[X]$-sentence.
\end{enumerate*}
A sentence $\act(t_1,t_2)$ is called a \emph{transition rule}, which can also be written  in infix notation $t_1 \mathrel{\act} t_2$.
An \emph{atomic sentence} is a ground equation  $t=t'$ or a ground transition rule of the form $t_1\mathrel{\varrho}t_2$, where $\varrho\in L$ is a transition label.

Besides the above core connectives, we also make use of the following convenient (and standard) abbreviations:
$\land\Phi\coloneqq\neg\lor_{\phi\in\Phi}\neg\phi$ for finite conjunctions;
$\bot\coloneqq\lor\emptyset$ for falsity;
$\top\coloneqq\land\emptyset=\neg\bot$ for truth;
$\phi_1\Rightarrow \phi_2\coloneqq \neg\phi_1\vee\phi_2$ for implications; and
\(\Forall{X} \phi' \coloneqq \neg \Exists{X} \neg \phi'\) for universally quantified sentences.

\begin{remark}
Any signature morphism $\chi\colon\Sigma\to\Sigma'$ can be canonically extended to a \emph{sentence-translation function} $\chi\colon\Sen^\TA(\Sigma)\to\Sen^\TA(\Sigma')$ given by:
\begin{itemize}[topsep=3pt]
\item $\chi(t = t') \coloneqq (\chi(t)=\chi(t'))$; 
\item $\chi(t_1 \mathrel{\act} t_2) \coloneqq \chi(t_1) \mathrel{\chi(\act)} \chi(t_2)$;
\item $\chi(\neg\phi) \coloneqq \neg\chi(\phi)$; 
\item $\chi(\lor\Phi) \coloneqq \lor\chi(\Phi)$; and
\item $\chi(\Exists{X}\phi') \coloneqq \Exists{X'}\chi'(\phi')$, where \(X' = \{ \pos{v_i,\chi(s),\Sigma'} \mid \pos{v_i,s,\Sigma} \in X \}\) and $\chi':\Sigma[X]\to\Sigma'[X']$ is the extension of $\chi:\Sigma\to\Sigma'$ mapping each $\pos{v_i,s,\Sigma}\in X$ to $\pos{v_i,\chi(s),\Sigma'}\in X'$.

\end{itemize}
Moreover, this sentence-translation mapping is functorial in \(\chi\).
\end{remark}

For the sake of simplicity, we identify variables only by their name and sort, provided that there is no danger of confusion.
Using this convention, each inclusion morphism $\iota\colon\Sigma\hookrightarrow\Sigma'$ determines an inclusion function $\iota\colon\Sen^\TA(\Sigma)\hookrightarrow\Sen^\TA(\Sigma')$, which corresponds to the approach of classical model theory.
This simplifies the presentation greatly. 
A situation when we cannot apply this convention arises when translating a $\Sigma$-sentence $\Exists{X}\phi$ along the inclusion $\iota_X:\Sigma\hookrightarrow\Sigma[X]$.

\subparagraph*{Satisfaction relation}
Actions are interpreted as binary relations in models.
Given a model $\A$ over a signature $\Sigma$, and actions $\act,\act_1,\act_2\in A$, we have:
\begin{itemize}[topsep=3pt]
\item $(\act_1\comp\act_2)^\A=\act_1^\A \comp \act_2^\A$ (i.e., diagrammatic composition of binary relations);
\item $(\act_1\cup\act_2)^\A=\act_1^\A \cup \act_2^\A$ (the union of binary relations); and
\item $(\act^*)^\A= (\act^\A)^*$ (the reflexive and transitive closure of binary relations).
\end{itemize}
We define the \emph{satisfaction relation} between models and sentences as follows:
\begin{itemize}[topsep=3pt]
\item $\A\models t=t'$ iff $t^\A=t'^\A$;
\item $\A\models t_1 \mathrel{\act} t_2$ iff $(t_1^\A, t_2^\A) \in \act^\A$;
\item $\A\models \neg\phi$ iff $\A\not\models\phi$;
\item $\A\models\lor\Phi$ iff $\A\models\phi$ for some sentence $\phi\in \Phi$; and
\item $\A\models\Exists{X}\phi'$ iff $\A'\models\phi'$ for some expansion $\A'$ of $\A$ to the signature $\Sigma[X]$.
\end{itemize}
%%%
\begin{proposition} [Satisfaction condition]  \label{prop:sat-cond} 
For all signature morphisms $\chi:\Sigma\to\Sigma'$,
all $\Sigma'$-models $\A$ and
all $\Sigma$-sentences $\phi$
we have:
$\A\red_{\chi}\models\phi
  \quad\text{iff}\quad
  \A\models\chi(\phi)$.
\end{proposition}
The satisfaction condition for $\TA$ is given in \cite{go-icalp24}.
%%% %%% %%% %%% %%% %%% %%% %%% %%% 
\subparagraph*{Logical framework}
%%% %%% %%% %%% %%% %%% %%% %%% %%% 
Throughout this paper, we work within a fragment $\frag$ of $\TA$ which is obtained by 
\begin{enumerate*}[label=(\alph*)]
\item restricting the category of signatures -- without prohibiting signature extensions with constants -- and 
\item discarding a subset of sentence or/and action operators from the grammar which is used to define sentences in $\TA$.
\end{enumerate*}
When there is no danger of confusion we drop
\begin{enumerate*}[label=(\alph*)] 
\item the subscript $\frag$ from the notations $\Sig^\frag$ and $\Sen^\frag$, and
\item the subscript $\TA$ from the notation $\Mod^\TA$.
\end{enumerate*}
We denote by $\Sen_0:\Sig\to \Set$ the sub-functor of $\Sen:\Sig\to \Set$, 
which maps each signature $\Sigma$ in $\frag$ to the set $\Sen_0(\Sigma)$ of atomic sentences from $\Sen(\Sigma)$.
%%% %%% %%% %%% %%% %%% %%% %%% %%%
\section{Related concepts}
%%% %%% %%% %%% %%% %%% %%% %%% %%%
We define the necessary concepts and present the existing results for presenting our advancements.
We make the following notational conventions:
\begin{itemize}[topsep=3pt]
\item Recall that in any category $\mathcal{C}$, the notation $|\mathcal{C}|$ refers to the class of objects in $\mathcal{C}$.

\item Let $\Class$ denote the large category whose objects are classes and whose morphisms are class functions.
\footnote{By contrast, a small category is one in which both the collection of objects and the collection of morphisms form sets, rather than proper classes.}

\item Assume a set $C$ and a cardinal $\alpha$. 
Let $\P_\alpha(C)\coloneqq\{C_1\subseteq C\mid  \card(C_1)<\alpha\}$, the set of all subsets of $C$ of cardinality strictly less than $\alpha$.
We write $C_1\subseteq_\alpha C$ if $C_1\in\P_\alpha(C)$.
\end{itemize}
%
%%% %%% %%% %%% %%% %%% %%% %%% %%% 
\subparagraph*{Presentations}
%%% %%% %%% %%% %%% %%% %%% %%% %%% 
Let $\Sigma$ be a signature, and
assume a class of $\Sigma$-models $\mathbb{M}$ and 
a set of $\Sigma$-sentences~$\Phi$.
\begin{itemize}[topsep=3pt]
\item We write $\mathbb{M}\models\Phi$ if $\A\models\Phi$ for all models $\A\in\mathbb{M}$.
\item Let $\mathbb{M}^\bullet=\{\varphi\in\Sen(\Sigma)\mid \mathbb{M}\models \Phi \}$, the set of sentences satisfied by all models in $\mathbb{M}$.
\item Let $\Phi^\bullet=\{\A\in|\Mod(\Sigma)| \mid \A\models \Phi\}$, the class of models which satisfy $\Phi$.
\item Let $\Mod(\Sigma,\Phi)$ be the full subcategory of $\Mod(\Sigma)$ of models which satisfy $\Phi$.
\end{itemize}
A \emph{presentation} is a pair $(\Sigma,\Phi)$ consisting of a signature $\Sigma$ and a set of $\Sigma$-sentences $\Phi$.
A \emph{theory} is a presentation $(\Sigma,\Phi)$ such that $\Phi=\Phi^{\bullet\bullet}$.
A \emph{presentation morphism} $\chi:(\Sigma,\Phi)\to(\Sigma',\Phi')$ consist of a signature morphism $\chi:\Sigma\to \Sigma'$ such that $\Phi'\models_{\Sigma'}\chi(\Phi)$.
A \emph{theory morphism} is a presentation morphism between theories.
%%% %%% %%% %%% %%% %%% %%% %%% %%% 
\subparagraph*{Substitutions}
%%% %%% %%% %%% %%% %%% %%% %%% %%% 
Let $\Sigma$ be a signature, 
$C_1$ and $C_2$ two $S$-sorted sets of new constants for $\Sigma$ .
A substitution $\theta : C_1 \to C_2$ over $\Sigma$ is a mapping from $C_1$ to $T_\Sigma(C_2)$.
As in case of signature morphisms, a substitution $\theta : C_1 \to C_2$ determines 
\begin{itemize}[topsep=3pt] 
\item a sentence functor $\Sen(\theta):\Sen(\Sigma[C_1])\to\Sen(\Sigma[C_2])$, which preserves $\Sigma$ and maps each constant $c\in C_1$ to a term $\theta(c)$, and 
\item a reduct functor $\red_\theta:\Mod(\Sigma[C_2])\to\Mod(\Sigma[C_1])$, 
which preserves the interpretation of $\Sigma$ and 
assigns to each constant $c\in C_1$ the interpretation of the term $\theta(c)\in T_\Sigma(C_2)$ in the category $\Mod(\Sigma[C_2])$, that is, for all $\A\in|\Mod(\Sigma[C_2])|$ we have 
\begin{enumerate*}[label=(\alph*)]
\item $(\A\red_\theta)_s=\A_s$ for all sorts $s$ in $\Sigma$,
\item $x^{(\A\red_\theta)}=x^\A$ for all function or transition symbols $x$ in $\Sigma$, and
\item $c^{(\A\red_\theta)}=\theta(c)^\A$ for all constants $c\in C_1$.
\end{enumerate*} 
\end{itemize}
As in the case of signature morphisms, we use $\theta$ to denote both the substitution $\theta:C_1\to C_2$ and the functor $\Sen(\theta):\Sen(\Sigma[C_1])\to\Sen(\Sigma[C_2])$.
The following result is a straightforward generalization of \cite[Proposition 5.6]{Diaconescu2008}.
%%%
\begin{proposition} [Satisfaction condition for substitutions]  \label{prop:sat-cond-subst} 
For all substitutions $\theta:C_1\to C_2$,
all $\Sigma[C_2]$-models $\A$ and
all $\Sigma[C_1]$-sentences $\phi$
we have:
$\A\red_{\theta}\models\phi
  \quad\text{iff}\quad
  \A\models\theta(\phi)$.
\end{proposition}
A \emph{reachable (transition) algebra}~\cite{Petria07} defined over a signature $\Sigma$ is a (transition) $\Sigma$-algebra $\A$ such that the unique homomorphism $h:T_\Sigma\to \A$ is surjective.
By \cite[Proposition 8]{gai-cbl}, a $\Sigma$-algebra $\A$ is reachable iff for all sets of new constants $C$ for $\Sigma$ and all expansions $\B$ of $\A$ to $\Sigma[C]$, 
there exists a substitution $\theta:C\to\emptyset$ such that $\A\red_\theta=\B$.

%%% %%% %%% %%% %%% %%% %%% %%% %%% 
\section{Compactness and Upward L\"owenheim-Skolem Property} \label{sec:compact}
%%% %%% %%% %%% %%% %%% %%% %%% %%% 
$\TA$ is very expressive and its sentences can control the cardinality of all its models under the the following assumptions:
\begin{itemize}[topsep=3pt]
\item Generalized Continuum Hypothesis (GCH) holds, that is, 
for any infinite cardinal number $\alpha$, the next larger cardinal is exactly $2^\alpha$; and
\item there are no inaccessible cardinals ($\neg\mathrm{IC}$).
\end{itemize} 
\begin{theorem}[Categoricity] \label{th:expansion}
Assume that $\frag=\TA$.
Let $\A$ be a model defined over a signature  $\Sigma=(S,F,L)$.
There exist a sort preserving inclusion of signatures $\iota^\circ:\Sigma\hookrightarrow \Sigma^\circ$ and a reachable $\iota^\circ$-expansion $\A^\circ$ of $\A$ such that any model of $(\Sigma^\circ,Th(\A^\circ))$ is isomorphic to $\A^\circ$, where $Th(\A^\circ)=\{\phi\in \Sen(\Sigma^\circ) \mid \A^\circ\models\phi\}$.
\end{theorem}
Theorem~\ref{th:expansion} is proven assuming $\mathrm{ZFC} + \mathrm{GCH} + \neg\mathrm{IC}$. These axioms are consistent relative to $\mathrm{ZFC}$—that is, if $\mathrm{ZFC}$ is consistent, then so is the extended theory. Consequently, Theorem~\ref{th:expansion} (and its conclusion) cannot be refuted within $\mathrm{ZFC}$. This illustrates that certain logical properties cannot be established within $\TA$ from ZFC alone.
The following example shows that joint Robinson consistency property fails in $\TA$.
\begin{example}[Robinson Consistency]
Consider a complete (first-order) theory $(\Sigma,T)$ with at least two non-isomorphic models $\A$ and $\B$.
Construct a pushout as illustrated on the left side of the diagram below.
\begin{center}\small
\begin{tikzcd}
\Sigma^\circ \ar[r,"\chi_2"] & \Sigma' \\
\Sigma \ar[r,"\iota^\circ",hook] \ar[u,"\iota^\circ",hook]& \Sigma^\circ \ar[u,"\chi_1"{right}]
\end{tikzcd}
\hspace{1cm}
\begin{tikzcd}
(\Sigma^\circ,Th(\B^\circ)) \ar[r,"\chi_2"] & (\Sigma', \chi_1(Th(\A^\circ)) \cup \chi_2(Th(\B^\circ)) ) \\
(\Sigma,T) \ar[r,"\iota^\circ",hook] \ar[u,"\iota^\circ",hook]& (\Sigma^\circ, Th(\A^\circ)) \ar[u,"\chi_1"{right}]
\end{tikzcd}
\end{center}
Since $\A\models T$ and $\A^\circ\red_\Sigma=\A$, we have $T\subseteq Th(\A^\circ)$.
Similarly, since $\B\models T$ and $\B^\circ\red_\Sigma=\B$, we have $T\subseteq Th(\B^\circ)$.
However, $\chi_1(Th(\A^\circ)) \cup \chi_2(Th(\B^\circ))$ is not consistent, because any model of $\chi_1(Th(\A^\circ)) \cup \chi_2(Th(\B^\circ))$, by Theorem~\ref{th:expansion}, would imply that $\A$ is isomorphic to $\B$.
\end{example}
Typically, the cardinality of a model $\mathcal{A}$ over a signature $\Sigma=(S,F,L)$ is defined as the sum of the cardinalities of its carrier sets, that is, $\card(\A)=\sum_{s\in S}\card(\A_s)$.
The following example illustrates that the classical Upward L\"owenheim-Skolem (ULS) property does not hold when sorts are treated as unary predicates rather than as distinct domains. 
\begin{example}
Let $\Sigma$ be a signature with countably infinitely many sorts $\{s_n\}_{n<\omega}$, where each sort $s_n$ has exactly one constant symbol $c_n : \to s_n$, and there are no transition labels.
Define the set of equations $\Gamma \coloneqq \{\Forall{x_n} (x_n = c_n) \mid n<\omega\}$, which asserts that each sort $s_n$ contains exactly one element.
\end{example}
Note that $\Sigma$ is a first-order signature, and $\Gamma$ is a set of $\Sigma$-sentences that has an infinite model. However, $\Gamma$ has no models of cardinality greater than $\omega$, indicating that the ULS property, as traditionally stated, fails to apply in this setting.
The following definition of the ULS property applies to many-sorted first-order logic.

\begin{definition}[Upward L\"owenheim-Skolem Property]
 $\alpha$-\emph{Upward L\"owenheim-Skolem} property holds for an infinite cardinal $\alpha$ whenever
for  all signatures $\Sigma=(S,F,L)$, all $\Sigma$-models $\A$ and all sorts $s\in S$ such that $\alpha\geq\card(\Sen(\Sigma))$ and $\alpha \geq\card(\A_s)\geq \omega$ there exists an elementary extension $\B$ of $\A$ such that $\card(\B_s)=\alpha$. 
\end{definition}
Recall that \emph{$\alpha$-compactness} holds whenever
for all signatures $\Sigma$ and all sets of $\Sigma$-sentences $\Phi$,
$\Phi$ has a model 
iff all subsets $\Psi\in \P_\alpha(\Phi)$ have a model.
From Theorem~\ref{th:expansion}, one can show that both ULS property and compactness fail.
\begin{corollary}\label{cor:usl}
Assume that $\frag=\TA$.
Then
\begin{enumerate*}[label=(\alph*)]
\item $\alpha$-ULS fails for all cardinals $\alpha>\omega$, and 
\item $\alpha$-compactness fails for all infinite cardinals $\alpha$.
\end{enumerate*}
\end{corollary}
Corollary~\ref{cor:usl} shows that the iteration operator for actions provides enough expressive power to control the cardinality of models. Moreover, compactness fails in all its forms. In this setting, we aim to develop a model construction method that does not rely on compactness, yet still supports the proof of logical properties such as Downward L\"owenheim-Skolem (DLS) theorem and OTT. It is worth noting that the results presented in the following sections hold without assuming either GCH or $\lnot \mathrm{IC}$.
%%% %%% %%% %%% %%% %%% %%% %%% %%% 
\section{Forcing}\label{sec:forcing}
%%% %%% %%% %%% %%% %%% %%% %%% %%%
In this section, we present a forcing technique, originally proposed in \cite{go-icalp24}, which extends classical forcing from one signature to a category of signatures in a non-trivial way.
To illustrate the motivation behind our approach, we recall Example 28 from \cite{go-icalp24}.
%%%
\begin{example}
Let $\Sigma$ be a signature defined as follows:
\begin{enumerate*}[label=(\alph*)]
\item $S\coloneqq\{s_i\mid i\in \omega\}$,
\item $F\coloneqq\{c:\to s_0, \linebreak d:\to s_0\}$, and
\item $L\coloneqq\{\lambda\}$.
\end{enumerate*}
Let $\Phi$ be a set of $\Sigma$-sentences which consists of:
\begin{enumerate*}[label=(\alph*)]
\item $ \lambda^*(c, d)$, and
\item $(\Exists{x_n}\top) \Rightarrow \neg \lambda^n(c,d) $ for all $n\in\omega$, where $x_n$ is a variable of sort $s_n$.
\end{enumerate*}
\end{example}
The first sentence asserts that there is a transition from $c$ to $d$ in a finite number of steps.
For each natural number $n$, the sentence $(\Exists{x_n}\top) \Rightarrow \neg \lambda^n(c,d)$ states that if the sort $s_n$ is not empty then there is no transition from $c$ to $d$ in exactly $n$ steps. 
Classical model construction techniques typically involve introducing an infinite number of constants for each sort -- commonly referred to as \emph{Henkin constants} -- followed by the application of quantifier elimination methods that select witnesses for existentially quantified variables. 
Let $C=\{C_{s_n}\}_{n<\omega}$ be a collection of Henkin constants, that is, newly introduced constants such that each $C_{s_n}$ is countably infinite for all $n<\omega$.
Although the original theory $(\Sigma,\Phi)$ is satisfiable, its extension $(\Sigma[C],\Phi)$ becomes unsatisfiable upon the addition of these constants.
This discrepancy highlights the limitations of classical approaches and motivates the need for a novel method to handle the introduction of Henkin constants effectively.
%%% %% %%%
\begin{definition}[Forcing property] \label{def:forcing}
\emph{A forcing property} is a tuple $\mathbb P=(P,\leq,\Delta,f)$, where:
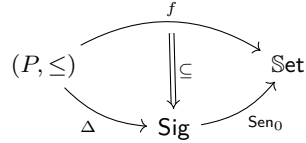
\begin{center}\small
\begin{tikzcd}[row sep=small]
(P,\leq) \ar[dr,bend right=20,"\Delta",swap] \ar[rr,bend left,"f", ""{below,name=U}] &  & \Set\\
& \Sig \ar[ur,bend right=20,"\Sen_0",swap]  \ar[Rightarrow,from = U,"\subseteq"] & 
\end{tikzcd}
\end{center}
\begin{enumerate}[topsep=3pt]
\item $(P,\leq)$ is a partially ordered set with a least element $0$.
		
The elements of $P$ are traditionally called conditions.
		
\item $\Delta: (P,\leq) \to \Sig$ is a functor, which maps each arrow $(p\leq q)\in (P,\leq)$ to an inclusion $\Delta(p)\subseteq \Delta(q)$.
\item $f:(P,\leq)\to\Set$ is a functor from the small category $(P,\leq)$ to the category of sets $\Set$  such that $f \subseteq \Delta;\Sen_0$ is a natural transformation, that is:
(\emph{a})~$f(p)\subseteq \Sen_0(\Delta(p))$ for all conditions $p\in P$, and
(\emph{b})~$f(p)\subseteq f(q)$ for all relations $(p\leq q)\in (P,\leq)$.
\item If $f(p)\models \phi$ then $\phi\in f(q)$ for some $q\geq p$, for all atomic sentences $\phi\in \Sen_0(\Delta(p))$.
\end{enumerate}
\end{definition}
A classical forcing property is a particular case of forcing property such that $\Delta(p)=\Delta(q)$ for all conditions $p,q\in P$.
As usual, forcing properties determine suitable relations between conditions and sentences.
\begin{definition} [Forcing relation]
Let $\mathbb P=\langle P,\leq,\Delta,f \rangle$ be a forcing property.
\emph{The forcing relation} $\Vdash$ between conditions $p\in P$ and sentences from $\Sen(\Delta(p))$ is defined by induction on the structure of sentences, as follows:
\begin{itemize}[topsep=3pt]
\item $p\Vdash \varphi$ if $\varphi \in f(p)$, for all atomic sentences $\varphi\in\Sen_0(\Delta(p))$.
		
\item  $p\Vdash (\act_1\comp\act_2)(t_1 , t_2)$ if $p\Vdash \act_1(t_1, t) $ and $p\Vdash \act_2(t ,t_2)$ for some $t\in T_{\Delta(p)}$.
		
\item $p\Vdash (\act_1\cup\act_2)(t_1,t_2)$ if $p\Vdash \act_1(t_1 ,t_2)$ or $p\Vdash \act_2(t_1,t_2)$.
		
\item $p\Vdash \act^*(t_1,t_2)$ if $p\Vdash \act^n(t_1,t_2)$ for some natural number $n<\omega$.
		
\item $p\Vdash \neg \phi$ if there is no $q\geq p$ such that $q\Vdash \phi$. 
		
\item $p\Vdash \vee \Phi$ if $p\Vdash \phi$ for some $\phi\in \Phi$.
		
\item $p\Vdash \Exists{X}\phi$ if $p\Vdash \theta(\phi)$ for some substitution $\theta:X \to T_{\Delta(p)}$.
\end{itemize}
The relation $p\Vdash \phi$ in $\mathbb{P}$, is read as $p$ forces $\phi$.
% We say that $p$ weakly forces $\phi$, in symbols, $p\Vdash^w \phi$, if $p\Vdash\neg\neg\phi$.
\end{definition}
%%% %% %%%
A few basic properties of forcing are presented below.
%%% %% %%%
\begin{lemma}[Forcing properties \cite{go-icalp24}] \label{lemma:fp}
Let $\mathbb P=(P,\leq,\Delta,f)$ be a forcing property. 
For all conditions $p\in P$ and all sentences $\phi\in\Sen(\Delta(p))$ we have: 
\begin{enumerate}[topsep=3pt] 
\item \label{fp-1} $p\Vdash \neg\neg \phi$ iff for each $q\geq p$ there is a condition $r\geq q$ such that $r\Vdash \phi$.
		
\item \label{fp-2} If $p\leq q$ and $p\Vdash \phi$ then $q\Vdash \phi$.
		
\item  \label{fp-3} If $p\Vdash \phi$ then $p\Vdash \neg\neg \phi$.
		
\item \label{fp-4} We can not have both $p\Vdash \phi$ and $p\Vdash \neg \phi$.
\end{enumerate}
\end{lemma}
%%% %% %%%
The second property stated in the above lemma shows that the forcing relation is preserved along inclusions of conditions.
The fourth property shows that the forcing relation is consistent, that is, a condition cannot force all sentences.
The remaining conditions are about negation.
%%% %% %%%
\begin{definition}[Generic set] \label{def:gs}
Let $\mathbb P=(P,\leq,\Delta,f)$ be a forcing property.
A subset of conditions $G\subseteq P$ is generic if
\begin{enumerate}[topsep=3pt] 
\item\label{gs-1} $G$ is an ideal, that is:
\begin{enumerate*}[label=(\alph*)]
\item $G\neq\emptyset$,
\item for all $p\in G$ and all $q\leq p$ we have $q\in G$, and
\item for all $p,q\in G$ there exists $r\in G$ such that $p\leq r$ and $q\leq r$; 
\end{enumerate*}
and
\item\label{gs-2} for all conditions $p\in G$ and all sentences $\phi\in \Sen(\Delta(p))$ there exists a condition $q\in G$ such that $q\geq p$ and either $q\Vdash \phi$ or $q\Vdash \neg \phi$ holds.
\end{enumerate}
We write $G\Vdash \phi$ if $p\Vdash \phi$ for some $p\in G$.
\end{definition}
A generic set $G$ describes a reachable model which satisfies all sentences forced by the conditions in $G$.
\begin{remark}\label{rem:gs}
Since $\Delta:(G,\leq)\to \Sig$ is a directed diagram of signature inclusions, one can construct a co-limit  $\mu: \Delta \Rightarrow \Delta_G$ of the functor $\Delta:(G,\leq)\to \Sig$ such that $\mu_p:\Delta(p)\to\Delta_G$ is an inclusion for all $p\in G$.
\footnote{Note that $\Delta_G$ refers both to the vertex of the colimit of the diagram $\Delta$, and the constant functor from the poset $(G, \leq)$ to the category $\Sig$, which maps every object in $G$ to the signature $\Delta_G$.}
\end{remark}
The results which leads to DLS and OTT are developed over a signature~$\Delta_G$ such as the one described in Remark~\ref{rem:gs}.
%%%
\begin{definition} [Generic model]
Let $\mathbb P=(P,\leq,\Delta,f)$ be a forcing property and $G\subseteq P$ a generic set.
A model $\A$ defined over $\Delta_G$ is a \emph{generic model} for $G$ iff for every sentence $\phi\in \bigcup_{p\in G}\Sen(\Delta(p))$, we have $\A\models \phi \mbox{ iff } G\Vdash \phi$.
\end{definition}
%%%
The notion of generic model is the semantic counterpart of the definition of generic set.
The following result shows that every generic set has a generic model.
\begin{theorem}[Generic Model Theorem \cite{go-icalp24}] \label{th:gm}
Let $\mathbb P=(P,\leq,\Delta,f)$ be a forcing property and $G\subseteq P$ a generic set. 
Then there is a generic model $\A$ for $G$ which is reachable.
\end{theorem}
%%% %%% %%% %%% %%% %%% %%% %%% %%%
\section{Semantic Forcing} \label{sec:semantic-forcing}
%%% %%% %%% %%% %%% %%% %%% %%% %%%
As an example of forcing property, we introduce \emph{semantic forcing} and study its properties.
Let us fix an arbitrary signature $\Sigma$ and let $\alpha\coloneqq\card(\Sen(\Sigma))$ be the \emph{power} of $\Sigma$.
Consider an $\S$-sorted set of new constants $C=\{C_s\}_{s\in S}$ such that $\card(C_s)=\alpha$ for all sorts $s\in S$.
These constants, traditionally known as Henkin constants, are used in the construction of models.
Throughout this section we assume the following:
\begin{enumerate}[topsep=3pt] 
\item \label{assume1} a partially ordered set $(\K,\preceq)$ with a least element $0_\K$;
\item \label{assume2} a signature functor $\Omega:(\K,\preceq)\to \Sig$ which maps
\begin{enumerate*}[label=(\alph*)]
\item $0_\K$ to $\Sigma$,
\item each element $\kappa\in\K$ to a signature $\Sigma[C_\kappa]$, where $C_\kappa$ is a subset of $C$ with cardinality strictly less than $\alpha$, that is, $C_\kappa\in\P_\alpha(C)$, and
\item each relation $(\kappa\preceq \ell)\in (\K,\preceq)$ to an inclusion of signatures $\Sigma[C_\kappa]\subseteq \Sigma[C_\ell]$; 
\end{enumerate*}
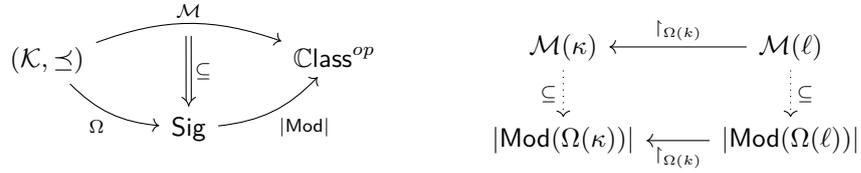
\begin{center}
\begin{tikzcd}[row sep=small]
(\K,\preceq) \ar[dr,bend right=20,"\Omega",swap] \ar[rr,bend left=20pt,"\M", ""{below,name=U}] &  & \Class^{op}\\
& \Sig \ar[ur,bend right=20,"|\Mod|",swap]  \ar[Rightarrow,from = U,"\subseteq"] & 
\end{tikzcd}
\hspace{1cm}
\begin{tikzcd}
\M(\kappa) \ar[d,dotted,"\subseteq"{left}] & \ar[l,"\red_{\Omega(k)}"{above}]\M(\ell) \ar[d,dotted,"\subseteq"]\\
{|\Mod(\Omega(\kappa))|} & {|\Mod(\Omega(\ell))|} \ar[l,"\red_{\Omega(k)}"]
\end{tikzcd}
\end{center}
\item \label{assume3} a model functor $\M:(\K,\preceq)\to \Class^{op}$ such that the following hold:
\footnote{$\Class^{op}$ is the opposite category of $\Class$, which means:
\begin{enumerate*}[label=(\alph*)]
\item it has the same objects as $\Class$, that is, classes, but
\item morphisms are reversed, that is, a morphism $f:A\to B$ in $\Class^{op}$ corresponds to a class function $f:B\to A$ in $\Class$.
\end{enumerate*}} 
\begin{enumerate}[topsep=3pt]
\item $\M\subseteq \Omega\comp|\Mod|$ is a natural transformation, that is,
\begin{enumerate*}[label=(\alph*)]
\item \label{assume3-a} $\M(\kappa)\subseteq |\Mod(\Omega(\kappa))|$ for all elements $\kappa\in \K$, and
\item \label{assume3-b} $\M(\ell)\red_{\Omega(\kappa)}\subseteq\M(\kappa)$ for all relations $(\kappa\preceq \ell)\in(\K,\preceq)$;
\end{enumerate*}  
\item Assume an element $\kappa\in \K$, a model $\A\in\M(\kappa)$ and a set of constants $C_1\subseteq_\alpha C\setminus C_\kappa$.
For all expansions $\B$ of $\A$ to $\Sigma[C_\kappa\cup C_1]$, there exists $\ell\succeq\kappa$ such that $C_\ell=C_\kappa\cup C_1$ and $\B\in\M(\ell)$.
\begin{center}
\begin{tikzcd}[row sep=small]
\M(\kappa)\ni\A \ar[r,dotted,no head]& \Sigma[C_\kappa] \ar[r,hook] & \Sigma[C_\kappa\cup C_1]=\Sigma[C_\ell] & \ar[l,dotted,no head]\B\in\M(\ell) \ar[lll,dotted,bend left=10,"\red_{\Sigma[C_\kappa]}"]
\end{tikzcd}
\end{center}
\end{enumerate}
\end{enumerate}
The condition \ref{assume3}\ref{assume3-b} implies that all sets of constants from $\P_\alpha(C)$ appear in some signature from the image of $\Omega$.
\begin{example} \label{ex:dls}
Let $\A$ be a model defined over a signature $\Sigma$.
\begin{enumerate}[topsep=3pt]
\item $\K=\{(\Sigma[C'],\A') \mid C'\in\P_\alpha(C) \text{ and }\A'\in |\Mod(\Sigma[C'])| \text{ such that }\A'\red_\Sigma=\A\}$, and

 $(\Sigma[C'],\A')\preceq(\Sigma[C''],\A'')$ iff $C'\subseteq C''$ and $\A''\red_{\Sigma[C']}=\A'$;
\item $\Omega:(\K,\preceq)\to \Sig$ is the forgetful functor mapping each $(\Sigma[C'],\A')$ to $\Sigma[C']$;
\item $\M:(\K,\preceq)\to \Class^{op}$ is the model functor which maps each $(\Sigma[C'],\A')\in \K$ to $\{\A'\}$.
\end{enumerate}
\end{example}
Example~\ref{ex:dls} shows that $(\K,\preceq)$ can have a more complex structure than $(\P_\alpha(C),\subseteq)$.
\begin{example} \label{ex:ott}
Let $\Phi$ be any set of sentences defined over a signature $\Sigma$.
\begin{enumerate}[topsep=3pt]
\item $\K=\{(\Sigma[C'],\Phi) \mid C'\in\P_\alpha(C)\}$, and
 $(\Sigma[C'],\Phi)\preceq(\Sigma[C''],\Phi)$ iff $C'\subseteq C''$;
\item $\Omega:(\K,\preceq)\to \Sig$ is the forgetful functor mapping each $(\Sigma[C'],\Phi)$ to $\Sigma[C']$;
\item $\M:(\K,\preceq)\to \Class^{op}$ is the functor which maps each $(\Sigma[C'],\Phi)\in \K$ to $|\Mod(\Sigma[C'],\Phi)|$,
the class of $\Sigma[C']$-models which satisfy $\Phi$.
\end{enumerate}
\end{example}
\begin{definition}[Semantic Forcing Property] \label{def:semantic-forcing}
The semantic forcing property $\mathbb{P}(\Omega,\M)=(P,\leq,\Delta,f)$ is defined as follows:
\begin{enumerate}[topsep=3pt]
\item A condition $p$ is a pair $(\kappa_p, \Phi_p)$,
where $\kappa_p\in \K$ and $\Phi_p\subseteq_\alpha \Sen(\Omega(\kappa_p))$ such that $\M(\kappa_p) \cap \Phi_p^\bullet\neq\emptyset$, that is, there exists at least one model in $\M(\kappa_p)$ which satisfies $\Phi_p$.
\item $p\leq q$ iff $\kappa_p\preceq \kappa_q$ and $\Phi_p\subseteq \Phi_q$, for all conditions $p=(\kappa_p, \Phi_p), q=(\kappa_q, \Phi_q)\in P$.
\item $\Delta:(P,\leq)\to \Sig$ is the functor mapping each condition $p=(\kappa_p,\Phi_p)\in P$ to $\Omega(\kappa_p)\in |\Sig|$.
\item Let $f(p)= \Sen_0(\Omega(\kappa_p)) \cap (\M(\kappa_p) \cap \Phi_p^\bullet)^\bullet$, 
for all conditions $p\in P$.
In other words,
$f(p)$ is the set of atomic sentences satisfied by all models of $\Phi_p$ from $\M(\kappa_p)$.
\end{enumerate}
\end{definition}
From a category theory perspective, several observations are worth noting.
\begin{remark} \label{rem:cat}
Definition~\ref{def:semantic-forcing} describe the following mathematical structures:
\begin{itemize}[topsep=3pt]
\item $\Lambda:(P,\leq)\to (\K,\prec)$, defined by $\Lambda(p)=\kappa_p$ for all $p\in P$, is a monotone function such that $\Delta=\Lambda\comp\Omega$.
\item $\Gamma:(P,\leq)\to \Set$, defined by $\Gamma(p)=\Phi_p$ for all $p\in P$, is a functor such that the inclusion $\subseteq ~: \Gamma\Rightarrow (\Lambda\comp\Omega\comp\Sen)$ is a natural transformation, that is:
\begin{enumerate}[topsep=3pt]
\item $\Gamma(p)\subseteq\Sen(\Delta(p))$ for all conditions $p\in P$, and
\item $\Gamma(p)\subseteq \Gamma(q)$ for all relations $(p\leq q)\in(P,\leq)$.
\end{enumerate}
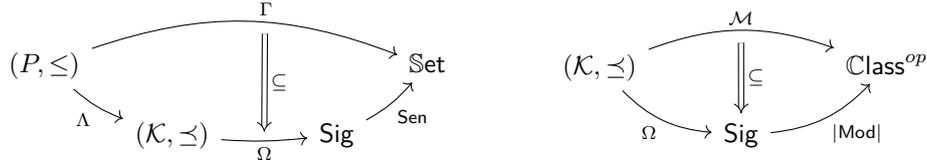
\begin{center}
\begin{tikzcd}[column sep=small,row sep=small]
(P,\leq) \ar[rrrr,bend left=20,"\Gamma"{above right=0pt and 3pt}, ""{below right= 0pt and 5pt,name=U}]  \ar[dr,bend right=10,"\Lambda",swap]& & & & \Set\\
& (\K,\preceq) \ar[rr,bend right=5,"\Omega",swap]& \ar[Rightarrow,from = U,"\subseteq"] & \Sig \ar[ur,bend right=10,"\Sen",swap] & 
\end{tikzcd}
\hspace{1cm}
\begin{tikzcd}[row sep=small]
(\K,\preceq) \ar[dr,bend right=20,"\Omega",swap] \ar[rr,bend left=20pt,"\M", ""{below,name=U}] &  & \Class^{op}\\
& \Sig \ar[ur,bend right=20,"|\Mod|",swap]  \ar[Rightarrow,from = U,"\subseteq"] & 
\end{tikzcd}
\end{center}
\end{itemize}
\end{remark}
Let $p\in P$ be a condition.
\begin{enumerate}[topsep=3pt]
\item We let $\Mod(p)$ denote $\M(\Lambda(p))\cap \Gamma(p)^\bullet$, the class of models in $\M(\Lambda(p))$ satisfying $\Gamma(p)$.
\item We write $p\models \phi$ iff $\phi$ is satisfied by all models in $\Mod(p)$, in symbols, $\Mod(p)\models \phi$. 
\end{enumerate}
The perspective of category theory can be disregarded if the reader is not familiar with it.
However, Remark~\ref{rem:cat} lays the groundwork for a generalization to an abstract level, as defined by the notion of institution~\cite{gog-ins}.

We define a notion of distance between conditions, which provides a principled method for regulating the addition of Henkin constants from $C$ to the starting signature $\Sigma$.

\begin{definition}[Distance \& weak forcing]
The distance between two conditions $p$ and $q$, with $p\leq q$, is defined as the number of constants and sentences that must be added to $p$ to obtain $q$, that is, $d(p,q):=\card(\Delta(q)\setminus\Delta(p))+\mathrm{card}(\Gamma(q)\setminus\Gamma(p))$.
A condition $p$ weakly forces $\phi$, in symbols, $p\Vdash^w \phi$, 
if for all $q\geq p$ there exists $r\geq q$ such that $d(q,r)< \omega$ and $r\Vdash \phi$.
\end{definition}
Using Lemma~\ref{lemma:fp}~(\ref{fp-2}) and the reflexivity of $\leq$, 
it is not difficult to show that $p\Vdash \phi$ implies $p\Vdash^w \phi$.
In the classical setting, weak forcing does not involve any notion of distance:
$p\Vdash^w \phi$ if and only if, 
for all $q\geq p$, there exists $r\geq q$ such that $r\Vdash \phi$; and this, in turn, holds if and only if $p\Vdash \neg\neg\phi$.
The notion of distance is introduced in this paper to extend the applicability of the results to all signatures -- including those of singular cardinality -- by controlling the number of constants introduced during the construction of the generic sets.
\begin{theorem} [Semantic Forcing Theorem] \label{th:sfp}
For all conditions $p \in P$ and all sentences $\phi\in\Sen(\Delta(p))$, we have:
\begin{enumerate*}[label=(\alph*)]
\item\label{fst1} $p\models \phi$ ~~~iff~~~
\item\label{fst2} $p\Vdash^{w} \phi$
\end{enumerate*}
\end{theorem}
%
%%% %%% %%% %%% %%% %%% %%% %%% %%%
\section{Downward L\"owenheim-Skolem Theorem}\label{sec:DLS}
Consider a semantic forcing property $\mathbb{P}(\Omega,\M)$ such that the functors $\Omega$ and $\M$ are defined as in Example~\ref{ex:dls}.
We use $\mathbb{P}(\Omega,\M)$ to give a proof of Downward L\"owenheim-Skolem Theorem.
The key to proving any result based on forcing is the construction of a generic set for a given condition $p$.
\begin{lemma}[Existence]\label{lemma:DLS}
Let  $\mathbb{P}(\Omega,\M)=(P,\leq,\Delta,f)$ be a semantic forcing property such that the functors $\Omega$ and $\M$ are defined as in Example~\ref{ex:dls}.
Any condition $p\in P$ belongs to a generic set $G_p$.
\end{lemma}
\begin{proof}
Let $pair:\alpha\times\alpha\to\alpha$ be any bijection such that for all ordinals $i,j,\beta<\alpha$ if $\beta=pair(i,j)$ then $i\leq\beta$.
For any condition $q\in P$, let $\gamma(q):\alpha\to \Sen(\Delta(q))$ be an enumeration of the sentences in $\Sen(\Delta(q))$.

First, we define an increasing chain of conditions $p_0\leq p_1\leq \ldots \leq p_\beta\leq \ldots$
with the following property: for all ordinals $\beta<\alpha$ there exists $n<\omega$ such that $d(p,p_{\beta})\leq n\cdot\mathrm{card}(\beta)$.\footnote{When $\alpha$ is a singular cardinal, the entire set of constants $C$ may be used along a chain $\{p_i\}_{i<\beta}$ with $\beta<\alpha$. 
To avoid this uncontrolled increase in the size, we use the distance measure $d$.}
We proceed by induction on ordinals $\beta<\alpha$.
\begin{description}
\item[($\beta=0$)] Let $p_0\coloneqq p$.
\item[($\beta\Rightarrow\beta+1$)]
Let $(i,j)\coloneqq pair^{-1}(\beta)$.
Notice that $i\leq\beta$, which means $p_i$ is already defined. 
Condition $p_{\beta+1}$ is obtained by adding a finite set of constants from $C$ and a finite set of sentences to $p_\beta$, according to a case distinction detailed below:
\begin{description}
\item[($q\Vdash \gamma(p_i,j)$ for some $q\geq p_\beta$)]
Since $q\Vdash \gamma(p_i,j)$, by Theorem~\ref{th:sfp}, $q\models \gamma(p_i,j)$.
It follows that $r\coloneqq (\Lambda(p_{\beta}),\Gamma(p_{\beta})\cup\{\gamma(p_i,j)\})\in P$. 
Since $r \vDash \gamma(p_i,j)$,
by Theorem~\ref{th:sfp}, there exists $p_{\beta+1}\geq r$ such that $p_{\beta+1}\Vdash\gamma(p_i,j)$ and $d(r,p_{\beta+1})<\omega$.
By induction hypothesis, there exists $n<\omega$ such that $d(p,p_{\beta+1})\leq n\cdot\card(\beta+1)$.
\item[($q\not\Vdash \gamma(p_i,j)$ for every $q\geq p_\beta$)] 
Let $p_{\beta+1}\coloneqq p_\beta$.
In this case, $p_{\beta+1}\Vdash \neg\gamma(p_i,j)$.
\end{description}
\item[($\beta<\alpha$ is a limit ordinal)]
At this point, for all $i<\beta$ we have defined $p_i=(\Lambda(p_i),\Gamma(p_i))$, 
where $\Lambda(p_i)=(\Delta(p_i),\A_i)$ and 
$\A_i$ is an expansion of $\A$ to $\Delta(p_i)$.
\begin{itemize}[topsep=3pt]
\item Let $C_{\Lambda(p_\beta)}\coloneqq\bigcup_{i<\beta}C_{\Lambda(p_i)}$ and 
$\Delta(p_\beta)=\Sigma[C_{\Lambda(p_\beta)}]$ and
$\Gamma(p_\beta)\coloneqq\bigcup_{i<\beta}\Gamma(p_i)$.
Since $\card(C_{\Lambda(p_i)})\leq\card(\beta)$ for all $i<\beta$,
we get $\card(C_{\Lambda(p_\beta)})\leq\card(\beta)<\alpha$.
Similarly, $\card(\Gamma(p_\beta))\leq\card(\beta)<\alpha$.
\item A model $\A_i\in \M(\Lambda(p_i))$, where $i<\beta$, can be regarded as a pair $(\A,g_i:C_{\Lambda(p_i)}\to \A)$, 
consisting of the model $\A$ and an interpretation of the constants in $C_{\Lambda(p_i)}$ into the elements of $\A$ given by the function $g_i$.
Define $g_\beta\coloneqq \bigcup_{i<\beta}g_i$, let $\A_\beta\coloneqq(\A,g_\beta)$ and $\Lambda(p_\beta)\coloneqq(\Delta(p_\beta),\A_\beta)$.
\item Let $p_\beta\coloneqq(\Lambda(p_\beta),\Gamma(p_\beta))$.
By its definition, $\A_\beta\red_{\Delta(p_i)}=\A_i$ for all $i<\beta$.
Since $\A_i\models\Gamma(p_i)$ for all $i<\beta$,
by satisfaction condition, 
$\A_\beta\models\Gamma(p_\beta)$.
Hence, $p_\beta$ is well-defined.
Since for all $i<\beta$ there exists $n_i\in\mathbb{N}$ such that $d(p,p_i)\leq n_i\cdot \card(i)$, 
we get $d(p,p_{\beta})\leq \card(\beta)$.
\end{itemize}
\end{description}
Secondly, one can easily show that $G_p\coloneqq\{q\in P\mid q\leq p_\beta \text{ for some } \beta < \alpha \}$ is a generic set.
\end{proof}
\begin{theorem}[Downward L\"owenheim-Skolem Theorem] \label{th:DLS}
Let $\A$ be a model over a signature $\Sigma=(S,F,L)$ of power $\alpha$ such that $\card(\A_s)\geq \alpha$ for some sort $s\in S$.
Then there exists an elementary submodel $\B$ of $\A$ such that $\card(\B_s)=\alpha$.
\end{theorem}
%%% %%% %%% %%% %%% %%% %%% %%% %%%
\section{Omitting Types Theorem}\label{sec:OTT}
%%% %%% %%% %%% %%% %%% %%% %%% %%%
Consider a semantic forcing property $\mathbb{P}(\Omega,\M)$ such that $\Omega$ and $\M$ are defined as in Example~\ref{ex:ott}.
We use $\mathbb{P}(\Omega,\M)$ to give a proof of Omitting Types Theorem in $\TA$.
\begin{definition}[Types]
A type for a signature $\Sigma$ of power $\alpha$ is a set of sentences $T$ defined over $\Sigma[X]$, where $X$ is a block of variables for $\Sigma$ such that $\alpha^{\card(X)}\leq\alpha$.
A $\Sigma$-model $\A$ \emph{realizes} $T$ if $\B\models T$ for some expansion $\B$ of $\A$ to $\Sigma[X]$.
$\A$ \emph{omits} $T$ if $\A$ does not realize $T$.
\end{definition}
For all infinite cardinals $\alpha$ we have $\alpha<\alpha^{\mathrm{cf(\alpha)}}$, where $\mathrm{cf}(\alpha)$ is the cofinality of $\alpha$.
Consequently, the condition $\alpha^{\card(X)}\leq\alpha$ implies that $\card(X)<\mathrm{cf}(\alpha)$. 
The converse also holds under the assumption of GCH.
The inequality $\alpha^{\card(X)}\leq\alpha$ ensures that the number of mappings from $X$ to the set of ground terms -- that is, substitutions $\theta:X\to \emptyset$ -- does not exceed $\alpha$.
Note that if $\alpha$ is the cardinality of real numbers and $X$ is countable then $\alpha^{\card(X)}\leq\alpha$ holds.
\begin{definition}[Isolated types] \label{def:isolate}
Let $\Sigma$ be a signature of power $\alpha$.
A set of sentences $\Phi\subseteq\Sen(\Sigma)$ is said to \emph{isolate} a type $T\subseteq\Sen(\Sigma[X])$ iff there exist:
\begin{itemize}[topsep=3pt]
\item an $S$-sorted set of new constants $D$ for $\Sigma$ with $\card(D)<\alpha$,
\item a set $\Gamma \subseteq  \Sen(\Sigma[D])$ with $\card(\Gamma)<\alpha$ and such that $T\cup \Gamma$ is satisfiable over $\Sigma[D]$, and
\item a substitution $\theta:X\to D$,
\end{itemize}
such that  $\Phi\cup\Gamma\models_{\Sigma[D]}\theta(T)$.
We say that $\Phi$ \emph{locally omits} $T$ if $\Phi$ does not isolate $T$.
\end{definition}
Our definitions for type, realization, and isolation align with those provided in \cite{GainaBK23},
with one key difference: we do not restrict the block of variables $X$ to be finite.
Definition~\ref{def:isolate} is similar to the definition of locally omitting types for first-order logic without equality from \cite{keisler01}.
If the fragment $\frag$ of $\TA$ is closed under Boolean operators and first-order quantifiers, the above definition coincides with the classical one for isolated types.
\begin{lemma} \label{lemma:otp}
Assume that 
\begin{enumerate*}[label=(\alph*)]
\item $\frag$ is  semantically closed under Boolean operators and first-order quantifiers, and
\item $\frag$ is compact or $\alpha=\omega$.
\end{enumerate*}
Then $\Phi\subseteq \Sen(\Sigma)$ isolates a type $T\subseteq\Sen(\Sigma[X])$ iff there exists a set of sentences $\Gamma\in \P_\alpha(\Sen(\Sigma[X]))$ such that $T\cup\Gamma$ is satisfiable over $\Sigma[X]$ and $\Phi\cup \Gamma\models_{\Sigma[X]} T$.
\end{lemma}
The proof is conceptually identical with the proof of \cite[Lemma 45]{GainaBK23}.
%%%%%%%%%%%%%%%%%%%%%%%%%%%%%%%%%%%%%%%
\begin{definition}[Omitting Types Property] \label{def:OTP}
Let $\alpha$ be an infinite cardinal.
The fragment $\frag$ has $\alpha$-Omitting Types Property ($\alpha$-OTP) whenever
\begin{enumerate}[topsep=3pt] 
\item for all signatures $\Sigma$ of power $\alpha$,
\item all satisfiable sets of sentences $\Phi\subseteq \Sen(\Delta)$, and
\item all families of types $\{T_i\subseteq \Sen(\Sigma[X_i]) \mid i<\alpha \}$,
\end{enumerate}
such that $\Phi$ locally omits $T_i$ for all $i<\alpha$,
there exists a $\Sigma$-model of $\Phi$ which omits $T_i$ for all $i<\alpha$.
If $\alpha=\omega$ then we say that $\frag$ has OTP rather than $\frag$ has $\omega$-OTP.
\end{definition}
%%%%%%%%%%%%%%%%%%%%%%%%%%%%%%%%%%%%%%%

%%%%%%%%%%%%%%%%%%%%%%%%%%%%%%%%%%%%%%%
\begin{theorem}[Omitting Types Theorem] \label{th:OTT}
Let $\Sigma$ be a signature of power $\alpha$.
If $\alpha>\omega$ then we assume that fragment $\frag$ is compact.
Then $\frag$ has $\alpha$-OTP.
\end{theorem}
%%%%%%%%%%%%%%%%%%%%%%%%%%%%%%%%%%%%%%%
Any fragment $\frag$ obtained from $\TA$ by 
\begin{enumerate*}[label=(\alph*)]
\item discarding a subset of sentence or/and action operators, and
\item restricting the category of signatures -- without prohibiting signature extensions with constants --
\end{enumerate*}
has $\omega$-OTP.
Any star-free fragment $\frag$ of $\TA$  is compact and therefore has $\alpha$-OTP for all cardinals cardinal $\alpha\geq \omega$.
In particular, many-sorted first-order logic has $\alpha$-OTP for all cardinals cardinal $\alpha\geq \omega$.

%%%%%%%%%%%%%%%%%%%%%%%%%%%%%%%%%%%%%%%
\begin{lemma}\label{lemma:inf}
Assume that $\frag$ is closed under Boolean operators and first-order quantifiers.
Let $\Sigma=(S,F)$ be a signature, where
 $S\coloneqq\{s_n  \mid n>0\}$ and $F\coloneqq \emptyset$.
\begin{itemize}[topsep=3pt]
\item Let $\phi_n\coloneqq \Exists{z_n}\top \Rightarrow \Exists{x_1,\dots,x_n} \bigwedge_{i\neq j} x_i\neq x_j$ be a $\Sigma$-sentence,
where $n>0$,
where $z_n$ is a variable of sort $s_n$, and 
$x_1,\dots,x_n$ are variables of sort $s_1$.
Note that $\phi_n$ states that if there exists an element of sort $s_n$ then there exists at least $n$ elements of sort $s_1$.
\item Let $T \coloneqq \{\Exists{x_1,\dots,x_n} \bigwedge_{i\neq j} x_i\neq x_j \land \bigwedge_{i=1}^n  y \neq x_i  \mid n > 0\}$ be a type with one variable $y$ of sort $s_1$, expressing that there are infinitely many elements of sort $s_1$.
\end{itemize}
Then $\Phi\coloneqq\{\phi_n \mid n>0\}$ locally omits $T$.
\end{lemma}
%%%%%%%%%%%%%%%%%%%%%%%%%%%%%%%%%%%%%%%
Let $\Sigma$, $\Phi$ and $T$ be as defined in Lemma~\ref{lemma:inf}.
Recall that the set of Henkin constants  is $C=\{C_{s_n}\}_{n<\omega}$, where $\card(C_{s_n})=\omega$ for all $n<\omega$. 
Note that $(\Sigma, \Phi)$ is a satisfiable first-order presentation that does not involve the transitive closure operator $*$, as it contains no relation symbols.
By Lemma~\ref{lemma:inf}, $\Phi$ locally omits $T$. 
By Theorem~\ref{th:OTT}, there exists a model $\A$ for $\Phi$ which omits $T$.
Since every model of $(\Sigma[C], \Phi)$ realizes $T$, the model $\A$ must be constructed over a proper subset of $C$, in contrast to the classical first-order logic approach, where results are developed over the entire $\Sigma[C]$. Consequently, the OTT as presented in \cite{keisler73, Chang:1990, Hodges:1997, gai-ott,GainaBK23} does not apply in this context. 
%%%
\begin{lemma} \label{lemma:unwanted}
Let $\Sigma=(S,F)$ be a signature such that $S\coloneqq \{s\}$ and $F\coloneqq\{c:\to s\mid c\in\mathbb{R}\}$, where $\mathbb{R}$ denotes the set of real numbers.
Let $T\coloneqq \{x_i\neq c \mid c:\to s\in F \text{ and } i<\omega\}$ be a type with a countably infinite number of variables from $X\coloneqq\{x_i\mid i<\omega\}$.
Then $\Phi$ locally omits $T$, where $\Phi$ is any countable and satisfiable set of $\Sigma$-sentences.
\end{lemma}
By Theorem~\ref{th:OTT}, there exists a model $\A$ of $\Phi$ which omits $T$. 
Consequently, the number of elements in $\A$ that are not denotations of real numbers -- considered unwanted elements -- is finite and may therefore be regarded as negligible.
Lemma~\ref{lemma:unwanted} demonstrates that types involving infinitely many variables are not difficult to construct and can naturally arise in applications.
%%%%%%%%%%%%%%%%%%%%%%%%%%%%%%%%%%%%%%%
\section{$\omega$-Completeness} \label{sec:complete}
%%%%%%%%%%%%%%%%%%%%%%%%%%%%%%%%%%%%%%%
Although a transition algebra may satisfy a given set of sentences, it may not be relevant to the intended semantics.
In many cases, formal methods practitioners are interested in the properties of a restricted class of transition algebras, such as:
\begin{enumerate*}[label=(\alph*)]
\item those reachable through a set of constructor operators, 
\item those with a finite number of elements, or
\item those satisfying both conditions.
\end{enumerate*}
Using Omitting Types Theorem we can extend the completeness result in \cite{go-icalp24} to fragments obtained from $\frag$ by restricting the semantics to the aforementioned transition algebras.
Concretely, given a sound and complete proof system for $\frag$, we introduce additional proof rules so that the resulting system remains sound and complete for fragments of $\frag$ obtained by restricting the class of transition algebras to those of interest.

\begin{definition}[Entailment relation] \label{def:entail} 
 An \emph{entailment relation} is a family of binary relations between sets of sentences $\vdash=\{\vdash_\Sigma\}_{\Sigma\in|\Sig^\frag|}$ with the following properties:
\begin{longtable}{l l}
$(Monotonicity)\Frac{\Phi_1\subseteq \Phi_2}{\Phi_2 \vdash \Phi_1}$
& $(Transitivity)\Frac{\Phi_1\vdash \Phi_2 \Space \Phi_2\vdash \Phi_3}{\Phi_1\vdash \Phi_3}$\\ \\
$(Union)\Frac{\Phi_1\vdash \varphi_2 \text{ for all }\phi_2\in\Phi_2}{\Phi_1\vdash\Phi_2}$ & 
$(Translation)\Frac{\Phi_1 \vdash_\Sigma \Phi_2}{\chi(\Phi_1) \vdash_{\Sigma'} \chi(\Phi_2)}$ where $\chi:\Sigma\to\Sigma'$ \\
\end{longtable}
An entailment relation $\vdash$ is sound (complete) if $\vdash\subseteq \models$ ($\models \subseteq \vdash$).
\end{definition}
Completeness fails when $\frag$ is closed under $*$ and includes uncountable signatures~\cite[Proposition 15]{go-icalp24}.
Therefore, throughout this section, 
we assume that the fragment $\frag$ 
has only countable signatures if it is closed under $*$. 
 Additionally, we assume that $\frag$ is semantically closed under Boolean operators.

%%% %%% %%%
\subparagraph*{Constructor-based algebras}
Many algebraic structures can be naturally described using a set of constructor operators.
For instance, numbers can be defined using a constant $zero$ and a unary function symbol $succ$,
while lists can be defined using a constant $nil$ and a binary function symbol $cons$.

%%% %%% %%%
\begin{definition} [Constructor-based algebras]
Let $\Sigma=(S,F,L)$ be a signature and 
$F^c\subseteq F$ a subset of constructor operators.
We let $\Sigma^c$ denote the signature of constructors $(S,F^c)$.
The constructors create a partition of the set of sorts $S$.
\begin{enumerate}[topsep=3pt]
\item A sort $s\in S$ that has a constructor -- i.e., there exists a constructor $(\sigma:w\to s)\in F^c$ -- is called \emph{constrained}.
We denote the set of all constrained sorts by $S^c$.
\item A sort that is not constrained is called \emph{loose}.
We denote the set of all loose sorts by $S^l$.
\end{enumerate}
A \emph{constructor-based transition algebra} $\A$ is a transition algebra for which there exist
\begin{enumerate*}[label=(\alph*)]
\item a set of loose variable $Y=\{Y_s\}_{s\in S^l}$ and
\item an expansion $\B$ of $\A$ to $\Sigma[Y]$,
\end{enumerate*}
such that $\B\red_{\Sigma^c[Y]}$ is reachable.
\end{definition}
\begin{example}[Lists]\label{ex:list}
Let $\Sigma=(S,F)$ be an algebraic signature such that
\begin{enumerate*}[label=(\alph*)]
\item $S=\{Elt,List\}$ and
\item $F=\{empty:\to List, \_;\_:List~Elt\to List, add:List~List\to List\}$.
\end{enumerate*}
Let $F^c=\{empty:\to List, \_;\_:List~Elt\to List\}$ be a set of constructors.
\begin{itemize}[topsep=3pt]
\item $Elt$ is a loose sort while $List$ is a constrained sort.
\item Let $\A$ be a constructor-based algebra obtained from $T_{\Sigma^c}(\omega)$ by interpreting $add$ as follows: 
\begin{enumerate*}[label=(\alph*)]
\item $add^\A(\ell,empty)=\ell$, and
\item $add^\A(\ell,\ell';n)=add^\A(\ell,\ell');n$ 
\end{enumerate*}
for all lists of natural numbers $\ell,\ell'\in\A_{List}$ and all natural numbers $n\in\A_{Elt}$.
\end{itemize}
\end{example}
By refining the syntax with a subset of constructor operators and restricting the semantics to constructor-based transition algebras, we obtain a new logic $\frag^c$ from $\frag$. This restriction in semantics alters the satisfaction relation $\models$ between sets of sentences: $\Phi \models^c \phi$ means that all constructor-based transition algebras of $\Phi$ are models of $\phi$. Consequently, there may exist non-restricted models of $\Phi$ that do not satisfy $\phi$.
In order to obtain an institution, we need to restrict the class of signature morphisms such that the reduct of a constructor-based algebra is again a constructor-based algebra~\cite{gai-cbl,DBLP:journals/tcs/Gaina13,DBLP:journals/tcs/BidoitHK03}.
So, the category of signatures of $\frag^c$ has 
\begin{enumerate}[topsep=3pt]
\item objects of the form $(S,F^c\subseteq F,L)$, where $(S,F,L)$ is a signature in $\frag$, and 
\item signature morphisms of the form $\chi:(S,F^c\subseteq F,L)\to (S',F'^c\subseteq F',L')$, where
\begin{enumerate*}[label=(\alph*)]
\item $\chi:(S,F,L)\to (S',F',L')$ is a signature morphism in $\frag$,
\item constructors are preserved, that is, $\chi(F^c)\subseteq F'^c$, and
\item constructors are reflected, that is, for all $\sigma':w'\to \chi(s)\in F'^c$ there exists $\sigma:w\to s\in F^c$ such that $\chi(\sigma:w\to s)=\sigma':w'\to\chi(s)$.
\end{enumerate*}
\end{enumerate}
%%% %%% %%%
\subparagraph{Finite algebras} 
%%% %%% %%%
Computer science applications often focus on properties that hold in finite structures. Therefore, we refine the fragment $\frag^c$ defined above to include constructor-based transition algebras, where certain distinguished sorts are interpreted as finite sets.
Let $\frag^f$ be a fragment of $\frag^c$ obtained by 
\begin{enumerate*}[label=(\alph*)]
\item refining the syntax with a subset of sorts of finite domains, and
\item by restricting constructor-based models such that the sorts for finite domains are interpreted as finite sets.
\end{enumerate*}
This means that the category of signatures in $\frag^f$ has 
\begin{enumerate}[topsep=3pt]
\item objects of the form $(S^f\subseteq S,F^c\subseteq F,L)$, where $(S,F^c\subseteq F,L)$ is a signature in $\frag^c$, and
\item arrows of the form $\chi:(S^f\subseteq S,F^c\subseteq F,L)\to (S'^f\subseteq S',F'^c\subseteq F',L')$ such that
\begin{enumerate*}[label=(\alph*)]
\item $\chi:(S,F^c\subseteq F,L)\to (S',F'^c\subseteq F',L')$ is a signature morphism in $\frag^c$, and
\item the finite sorts are preserved, that is, $\chi(S^f)\subseteq S'^f$.
\end{enumerate*}
\end{enumerate}

\begin{definition}
Let $\vdash$ be an entailment relation for $\frag$.
The entailment relation $\vdash^f$ for $\frag^f$ is the least entailment relation that includes $\vdash$ and it is closed under the following proof rules: 
\begin{enumerate}[topsep=3pt]
\item $(CB)~\Frac{\Phi\vdash^f \Forall{var(t)}\psi(t) \text{ for all }  t\in T_{\Sigma^c}(Y)}{\Phi \vdash^f \Forall{x}\psi(x)}$,
where 
\begin{enumerate*}[label=(\alph*)]
\item $x$ is a variable of constrained sort,
\item $Y$ is a $S^l$-sorted block of variables such that $\card(Y_s)=\omega$ for all loose sorts $s\in S^l$, and
\item $var(t)$ is the set of all variables occuring in $t$.
\end{enumerate*}
\item $(FN)~\Frac{\Phi \cup \{ \Gamma_{\overline{n}} \}\vdash^f \psi \text{ for all } \overline{n}\in\omega^\alpha }{\Phi \vdash^f \psi}$, where
\begin{enumerate}[topsep=3pt]
\item  $\alpha\coloneqq \card(S^f)$ and $\{s_i\}_{i<\alpha}$ is an enumeration of $S^f$,
\item $\Gamma_{\overline{n}}$ denotes the set of sentences $\{\gamma_{s_i,\overline{n}(i)}\mid i<\alpha\}$ for all tuples $\overline{n}\in \omega^\alpha$, where 
\begin{itemize}[topsep=3pt]
\item $\overline{n}(i)$ is the natural number in position $i$ from $\overline{n}$, for all tuples $\overline{n}\in\omega^\alpha$ and  all $i<\alpha$;
\item $\gamma_{s,n}$ denotes the sentence $\Forall{x_1,\dots,x_{n+1}} \lor_{i\neq j} x_i= x_j$,
for all natural numbers $n<\omega$, all sorts $s\in S^f$ and
all variables $x_1,\dots,x_{n+1}$ of sort $s$.
\end{itemize}
\end{enumerate}
Note that $\gamma_{s,n}$ states that the sort $s$ has at most $n$ elements.
If $n=0$ then $\gamma_{0,s}=\Forall{x_1}\vee\emptyset$, which is satisfied only by transition algebras whose carrier set for the sort $s$ is empty. 
\end{enumerate}
\end{definition}
$(CB)$ asserts that to prove the property $\psi$ for an arbitrary element $x$, it suffices to establish $\psi$ for all constructor terms $t$ in $T_{(S,F^c)}(Y)$.
$(CB)$ should be disregarded if the base logic $\frag$ is not closed under first-order quantification.
$(FN)$ states that the number of elements of any sort in $S^f$ is finite.
It is straightforward to show that $\vdash^f$ is sound for $\frag^f$ provided that $\vdash$ is sound for $\frag$.
Establishing completeness is generally more challenging, but it can be achieved with the help of the OTP.
%%% %%% %%%
\begin{theorem}[Completeness] \label{th:complete}
The entailment relation $\vdash^f$ is complete for $\frag^f$ if $\vdash$ is complete for $\frag$ and $\frag$ has OTP.
\end{theorem}
%%% %%% %%% %%% %%% %%% %%% %%% %%%
\section{Conclusions}
%%% %%% %%% %%% %%% %%% %%% %%% %%%
In this contribution, we focused on two main aspects of $\TA$. First, we examined its model-theoretic properties, including the L\"owenheim-Skolem and Omitting Types properties. Due to its increased expressivity, TA does not possess the Upward L\"owenheim-Skolem property. We then developed a semantic forcing property based on recent developments from \cite{go-icalp24}, which enabled us to prove the Downward L\"owenheim-Skolem Theorem and the Omitting Types Theorem. Given that the full expressivity of $\TA$ is not required in many practical case studies, our study was conducted on a fragment of TA that restricts the syntax while preserving the semantics.

Secondly, we explored applications of the Omitting Types Theorem to formal methods. Software engineers often seek a restricted class of models for a given theory, such as constructor-based and/or finite models. We extended the system of proof rules proposed in \cite{go-icalp24} with new rules that are sound for these models of interest, ensuring that completeness is maintained. This approach provides sound and complete proof rules for fragments of TA obtained by restricting the syntax and/or the semantics.

An interesting future research direction involves developing specification and verification methodologies based on the proof rules we have defined for the fragments of $\TA$. 
Additionally, we aim to propose a notion of Horn clauses for $\TA$, which could make specifications defined in $\TA$ executable through rewriting or narrowing.
%%% %%% %%% %%% %%% %%% %%% %%% %%%
\bibliographystyle{plainurl}
\bibliography{ta}
%%% %%% %%% %%% %%% %%% %%% %%% %%%
\appendix

\section{Proof of the results presented in Section~\ref{sec:compact}}
\begin{proof}[Proof of Theorem~\ref{th:expansion} (Categoricity)]
For simplicity, we assume that $S$ is a singleton $\{s\}$.
We consider the non-trivial case when  $\alpha\coloneqq\card(\A_s)$ is an infinite cardinal.

First, let $\Sigma^\circ \coloneqq (S,F^\circ,L^\circ)$, where
\begin{itemize}[topsep=3pt]
\item $F^\circ \coloneqq F \cup \{c_i:\to s \mid i< \alpha\} \cup \{ \sigma_\beta : s~s\to s \mid \beta\leq \alpha \text{ is an infinite successor cardinal}\}$ $~\cup~ \{ \delta_\beta: s \to s \mid \beta \leq\alpha \text{ is a limit cardinal}\}$; 
\item $L^\circ \coloneqq L\cup \{ \preceq, \suc \}$.
\end{itemize}
We have defined an inclusion of signatures $\iota^\circ:\Sigma\hookrightarrow\Sigma^\circ$.

Secondly, we define a $\iota^\circ$-expansion $\A^\circ$ of $\A$.
Without loss of generality, we assume that $\A_s=\alpha$.
For each uncountable cardinal $\beta\leq \alpha$, we define:
\begin{description}
\item [($\beta$ is a successor cardinal)]
In this case, we have $\beta=\lambda^+$ for some cardinal $\lambda$.

By GCH, there exists a bijective function $b_\beta:\beta\to 2^\lambda$.

\item [($\beta$ is a limit cardinal)] 
In this case, $\beta$ can be expressed as the supremum of a strictly increasing sequence of smaller cardinals, that is,
$\beta=sup\{\lambda_i \mid i< cf(\beta) \}$, where 
$cf(\beta)$ is the cofinality of $\beta$ and
$\{\lambda_i\}_{i< cf(\beta)}$ is an increasing sequence of cardinals strictly smaller than $\beta$.
Let $f_\beta: cf(\beta)\to \beta$ defined by $f_\beta(i)=\lambda_i$ for all $i<cf(\beta)$.
\end{description}
Let $\A^\circ$ be a $\iota^\circ$-expansion of $\A$ defined as follows:
\begin{itemize}[topsep=3pt]
\item $c_i^{\A^\circ}=i$ for all all ordinals $i<\alpha$.
\item 
$\sigma_{\lambda^+}^{\A^\circ}(i,j)=
\left\{ \begin{array}{l l}
b_{\lambda^+}(i)(j) & \text{if } i < \lambda^+ \text{ and } j <\lambda \\ 
2 & \text{otherwise}
\end{array}\right.$,
for all infinite cardinals $\lambda$.

Each $i<\lambda^+$ can be regarded as a function in $2^\lambda$ and for all $j<\lambda$, the ordinal $\sigma^{\A^\circ}_{\lambda^+}(i,j)$ is the results of the applying $i$ to $j$.
\item $\delta_\beta^{\A^\circ}(i)=
\left\{\begin{array}{l l}
f_\beta(i) & \text{if } i<cf(\beta)\\
2 & \text{otherwise}
\end{array}\right.$,
for all limit cardinals $\beta$.

In this case, $\beta$ is the supremum of $\{\delta_\beta^{\A^\circ}(i)\}_{i<cf(\beta)}$.
\item $\preceq^{\A^\circ}\coloneqq \{ (i,j) \mid \text{ for all ordinals } i\leq j <\alpha \}$.
\item $\suc^{\A^\circ}\coloneqq \{(i,i+1)\mid \text{for all natural numbers }i<\omega\}$.
\end{itemize}

Thirdly, we show that each model of $(\Sigma^\circ,Th(\A^\circ))$ is isomorphic to $\A^\circ$.
Since $\A^\circ$ is reachable by the constants in $\{c_i\}_{i<\alpha}$, that is, 
$\A\models \Forall{x} \lor\{x=c_i\mid i<\alpha\}$,
it suffices to show that each model of $(\Sigma^\circ,Th(\A^\circ))$ is reachable by the constants in $\{c_i\}_{i<\alpha}$.
Concretely, given a $\Sigma^\circ$-model $\B$, we show that the following property holds: 
\begin{equation}
\{c_i^\B\mid i< \beta \} = \{a\in \B_s \mid a \prec^\B c_\beta\} \text{ for all infinite cardinals }\beta\leq\alpha.
\end{equation}
Obviously, $\{c_i^\B\mid i< \beta \} \subseteq \{a\in \B_s \mid a \prec^\B c_\beta\}$.
Therefore, we focus on the other inclusion.
We proceed by induction on infinite cardinals $\beta\leq\alpha$.
\begin{description}
\item[($\beta=\omega$)]
Assume that $a\prec^\B c_\omega^\B$, that is, $a\preceq^\B c_\omega^\B$ and $a\neq c_\omega^\B$. 
Notice that $\A^\circ$ satisfies the following sentences:
\begin{enumerate}[label=(S\arabic*)]
\item $\Forall{x} x\prec c_\omega \Rightarrow c_0 \suc^* x$, which says that any element smaller than $c_\omega$ is obtained from $c_0$ by applying successor finitely many times;
\item  $c_i \suc c_{i+1}\land c_i \prec c_\omega$ for all $i<\omega$, which says than $c_{i+1}$ is the successor of $c_{i+1}$ and $c_i$ is strictly small than $c_\omega$; and
\item $\Forall{x} x\prec c_\omega \Rightarrow (\Exists{y} x \suc y \land \Forall{z} x \suc z\Rightarrow z=y)$,
which says that any element smaller than $c_\omega$ has a unique successor.
\end{enumerate}
By the first sentence, $c_0^\B (\suc^\B)^*a$, which implies $c_0^\B (\suc^\B)^k a$ for some $k<\omega$.
By the first conjunct of the second sentence, $c_0^\B (\suc^\B)^k c_k^\B$.  
By the second conjunct of the second sentence and the third sentence, $c_k=a$.
\item [($\beta=\lambda^+$ for some infinite cardinal $\lambda$)]
Assume that $a\prec^\B c_\beta^\B$.
Notice that $\A^\circ$ satisfies the following sentences:
\begin{enumerate}[label=(S\arabic*)]
\setcounter{enumi}{3}
\item \label{sen:S4}
$\Forall{x,y} x\prec c_\beta \land y\prec c_\lambda \Rightarrow \sigma(x,y)\prec c_2$;
\item \label{sen:S5}
$\Forall{x,y} x\prec c_\beta \land y\prec c_\beta \Rightarrow ((\Forall{z} z\prec c_\lambda \Rightarrow \sigma(x,z)=\sigma(y,z))\Rightarrow x=y)$,
which together with the above sentence says that any element smaller than $c_\beta$ can be regarded as a function from $\lambda$ to $2$; and
\item \label{sen:S6} $c_i\prec c_\beta$ for all $i < \beta$.
\end{enumerate}
Let $F:\lambda\to 2$ be the function defined by $F(i)=j$ iff $\sigma^\B_\beta(a,c_i^\B)=c_j^\B$ for all $i<\lambda$ and all $j<2$.
By the first two sentences above, $F$ is well-defined.
Let $k\coloneqq b_\beta^{-1}(F)$ and we show that $a=c_k^\B$:
\begin{enumerate}[label=(\alph*)]
\item Let $i<\lambda$ and $j<2$ such that $\sigma^\B_\beta(a,c_i^\B)=c_j^\B$.
We have $\sigma^{\A^\circ}_\beta(c_k^{\A^\circ},c_i^{\A^\circ}) = b_\beta(k)(i) = F(i)= j = c_j^{\A^\circ}$.
Since $\A^\circ\models \sigma_\beta(c_k,c_i)=c_j$, we get $\B\models\sigma_\beta(c_k,c_i)=c_j$.
It follows that $\sigma^\B_\beta(a,c_i^\B) = c_j^\B=\sigma^\B_\beta(c_k^\B,c_i^\B)$.
Hence, $\sigma^\B_\beta(a,c_i^\B)=\sigma^\B_\beta(c_k^\B,c_i^\B)$ for all $i<\lambda$.
\item By the induction hypothesis, $\sigma^\B_\beta(a,d)=\sigma^\B_\beta(c_k^\B,d)$ for all $d\prec^\B c_\lambda^\B$.
\item By the fifth sentence \ref{sen:S5}, we get $a=c_k^\B$.
\end{enumerate}
\item[($\beta\leq \alpha$ is a limit cardinal)]
Let $a\prec^\B c_\beta^\B$.
Notice that $\A^\circ$ satisfies the following sentences:
\begin{enumerate}[label=(S\arabic*)]
\setcounter{enumi}{6}
\item \label{sen:S7} $\Forall{x}x\prec c_\beta \Rightarrow \Exists{y} y \prec c_{cf(\beta)} \land x \prec \delta_\beta(y)$; and
\item \label{sen:S8} $\delta_\beta(c_i)=c_{f_\beta(i)}$ for all $i<cf(\beta)$.
\end{enumerate}
By the seventh sentence \ref{sen:S7}, 
there exists $d\prec^\B c_{cf(\beta)}^\B$ such that $a\prec^\B\delta_\beta^\B(d)$.
Since there are no inaccessible cardinals, $cf(\beta)<\beta$. 
By induction hypothesis, $d=c_i^\B$ for some $i<cf(\beta)$.
By the eighth sentence \ref{sen:S8}, $\delta_\beta^\B(c_i^\B)=c_{f_\beta(i)}^\B$.
It follows that $a\prec^\B c_{f_\beta(i)}^\B$, where $f_\beta(i)<\beta$. 
By induction hypothesis, $a=c_k^\B$ for some $k<f_\beta(i)$.
\end{description}
\end{proof}
%%% Corollary USL %%%
\begin{proof}[Proof of Corollary~\ref{cor:usl}]
Let $\Sigma=(S,F,L)$ be a signature such that $S=\{s\}$, $F=\emptyset$ and $L=\{\preceq\}$.
\begin{enumerate}
\item Let  $\beta<\alpha$ be two infinite cardinals.
Let $\A$ be a $\Sigma$-model such that $\A_s=\beta$ and $\preceq^\A=\leq$, the order among ordinals.
By Theorem~\ref{th:expansion}, there exist a sort preserving inclusion of signatures $\iota^\circ:\Sigma\hookrightarrow \Sigma^\circ$ and a $\iota^\circ$-expansion $\A^\circ$ of $\A$ such that all models of $Th(\A^\circ)$ are isomorphic with $\A^\circ$.
There is no model $\B$ of $Th(\A^\circ)$ such that $\card(\B_s)=\alpha$. 
Therefore, $\alpha$-ULS does not hold. 

\item Let $\alpha$ be an infinite cardinal.
Let $\A$ be a $\Sigma$-model such that $\A_s=\alpha$ and $\preceq^\A=\leq$, the order among ordinals.
By Theorem~\ref{th:expansion}, there exist an inclusion of signatures $\iota^\circ:\Sigma\hookrightarrow \Sigma^\circ$ and a $\iota^\circ$-expansion $\A^\circ$ of $\A$ such that all models of $Th(\A^\circ)$ are isomorphic with $\A^\circ$.
Let $\Sigma^\diamond$ be the signature obtained from $\Sigma^\circ$ by adding
\begin{enumerate*}[label=(\alph*)]
\item a new sort $st$,
\item a new set of constant symbols $\{ct_i:\to st\}_{i<\alpha}$, and
\item a function symbol $ft:st\to s$.
\end{enumerate*}
We define the following sets of $\Sigma^\diamond$-sentences:
\begin{enumerate}
\item $\Phi\coloneqq\{\Exists{y}\Forall{x} ft(x) \prec y \}\cup \{\Forall{x,y} ft(x)=ft(y)\Rightarrow x=y\}$, 
which says that $ft$ is bounded and injective;
\item $\Psi_\beta\coloneqq \{ ct_i \neq ct_j \mid i,j < \beta \text{ such that } i\neq j \}$ for all cardinals $\beta<\alpha$,
which implies that $\card(\B_{st})\geq \beta$ for any model $\B$ of $\Psi_\beta$; and
\item $T_\beta\coloneqq Th(\A^\circ) \cup \Phi\cup \Psi_\beta$.
\end{enumerate}
Let $\Psi_\alpha\coloneqq \bigcup_{\beta<\alpha} \Psi_\beta$ and $T_\alpha\coloneqq Th(\A^\circ) \cup \Phi\cup \Psi_\alpha$.
Notice that $\card(\B_{st})\geq \alpha$ for any model $\B$ of $\Psi_\alpha$.
It is not difficult to check that $T_\beta$ is satisfiable for all $\beta<\alpha$,
while $T_\alpha$ is not satisfiable:
\begin{itemize}[topsep=3pt]
\item[] Suppose towards a contradiction that there exists a model $\B$ of $T_\alpha$.
Since $\B\models \Psi_\alpha$, we have $\card(\B_{st})\geq\alpha$.
Since $\B\models Th(\A^\circ)$, $\card(\B_s)=\card(\A^\circ_s)=\alpha$.
Since $\B\models \Phi$, $ft^\B:\B_{st}\to \B_s$ is injective and bounded, which implies $\card(\B_{st}) < \card(\B_s)$.
Since $\card(\B_{st})\geq\alpha$ and $\card(\B_s)=\alpha$, we get $\alpha<\alpha$, which is a contradiction.
\end{itemize}
\end{enumerate}
\end{proof}
%%% %%% %%% %%% %%% %%% %%% %%% %%%
\section{Proofs of the results presented in Section~\ref{sec:semantic-forcing}}
%%% %%% %%% %%% %%% %%% %%% %%% %%%
\begin{lemma} \label{lemma:sfp}
Assume a condition $p\in P$,
an atomic sentence $\varphi\in\Sen_0(\Delta(p))$, 
a sentence $\phi\in\Sen(\Delta(p))$ and 
a set of sentences $\Phi\subseteq \Sen(\Delta(p))$.
\begin{enumerate}
\item \label{lemma:sfp-1} $p\models\varphi$ iff $f(p)\models \varphi$ iff $\varphi\in f(p)$.

\item \label{lemma:sfp-2} $p\models\phi$ implies $q\models \phi$, for all relations $(p\leq q)\in (P,\leq)$.

\item \label{lemma:sfp-3} $p\models \Phi$ iff
for all $q\geq p$ there exists $r\geq q$ such that $r\models\Phi$.
\end{enumerate}
\end{lemma}
\begin{proof}
We only prove the backwards implication of the third statement.
Suppose towards a contradiction that $p\not\models\Phi$.
Then $\A\models\neg\phi$ for some $\phi\in \Phi$ and some $\A\in\Mod(p)$.
Let $q\coloneqq (\Lambda(p),\Gamma(p)\cup\{\neg\phi\})$.
Notice that $q\geq p$ and for all $r\geq q$, we have $r\models \neg\phi$. 
It follows that for all $r\geq q$, we have $r\not\models\Phi$, which is a contradiction.
\end{proof}
First statement of  Lemma~\ref{lemma:sfp} show that a condition $p$ satisfies an atomic sentence $\varphi$ if and only if $\varphi$ is a semantic consequence of the atomic sentences of $p$.
The second statement establises that the ordering relation among conditions preserves the satisfaction of sentences.
The third statement asserts that a condition $p$ satisfies a set of sentences $\Phi$ if and only if $\Phi$ is satisfied by all conditions in all trees of conditions  rooted at $p$.
\begin{lemma} \label{lemma:up}
For any condition $p \in P$, the following hold:
\begin{enumerate}
\item  \label{lemma:up1} 
If $p\models (\act_1\comp\act_2)(t_1,t_2)$ then $q\models \act_1(t_1,c)$ and $q\models \act_2(c,t_2)$ 
for  some constant $c\in C$ and some condition $q \geq p$ such that $d(p,q)<\omega$.
\item  \label{lemma:up2} 
If $p\models (\act_1\cup\act_2)(t_1,t_2)$ then $q\models \act_1 (t_1,t_2)$ or $q\models \act_2(t_1,t_2)$
for some condition $q \geq p$ such that $d(p,q)<\omega$.
\item  \label{lemma:up3} 
If $p\models \act^*(t_1,t_2)$ then $q\models \act^n(t_1,t_2)$ 
for some natural number $n<\omega$ and some condition $q \geq p$ such that $d(p,q)<\omega$.
\item  \label{lemma:up4} 
If $p\models \vee\Phi$ then $q\models \phi$ 
for some sentence $\phi\in\Phi$ and some condition $q \geq p$ such that $d(p,q)<\omega$.
\item  \label{lemma:up5} 
If $p\models \Exists{X}\phi$ then $q\models\chi(\phi)$ 
for some bijective signature morphism $\chi:\Delta(p)[X]\to \Delta(q)$ which preserves $\Delta(p)$ and some condition $q \geq p$ such that $d(p,q)<\omega$.
\end{enumerate}
\end{lemma}
\begin{proof}
Let $p\in P$ be a condition.
\begin{enumerate}
\item Assume $p \models (\act_1\comp\act_2)(t_1,t_2)$.
Since $\Mod(p)\neq\emptyset$, there exists $\A\in \Mod(p)$.
It follows that $\A\models (\act_1\comp\act_2)(t_1,t_2)$. 
We have $(t_1^\A,d)\in \act_1^\A$ and $(d,t_2^\A)\in \act_2^\A$ for some $d\in \A_s$,
where $s$ is the sort of $t_1$ and $t_2$.
Let $c\in C\setminus C_{\Lambda(p)}$ be a constant of sort $s$.
Let $\B$ be an expansion of $\A$ to $\Delta(p)[c]$ which interprets $c$ as $d$. 
By our assumptions, there exists $\ell\succeq\Lambda(p)$ such that $C_\ell=C_{\Lambda(p)}\cup\{c\}$ and $\B\in \M(\ell)$.
Let $q\coloneqq (\ell,\Gamma(p)\cup\{\act_1(t_1,c),\act_2(c,t_2)\})$ and notice that $d(p,q)<\omega$.
By satisfaction condition, $\B\models \Gamma(p)$.
By definition, $(t_1^\B,d)\in \act_1^\B$ and $(d,t_2^\B)\in \act_2^\B$.
It follows that $\B\models \{\act_1(t_1,c),\act_2(c,t_2)\}$.
Therefore, $\B\in \Mod(q)$, which means that $q$ is well-defined.
\item Let $\A\in\Mod(p)$.
Since $p\models (\act_1\cup\act_2)(t_1,t_2)$, 
we have $\A\models (\act_1\cup\act_2)(t_1,t_2)$.
By semantics, $\A\models \act_i(t_1,t_2)$ for some $i\in \{1,2\}$.
Let $q\coloneqq (\Lambda(p),\Gamma(p)\cup\{\act_i(t_1,t_2)\})$ and notice that $d(p,q)<\omega$.
We have $\A\in\Mod(q)$, which means $q\in P$.
\item Let $\A\in\Mod(p)$.
Since $p\models \act^*(t_1,t_2)$, 
we have $\A\models \act^*(t_1,t_2)$.
By semantics, $\A\models \act^n(t_1,t_2)$ for some $n<\omega$.
Let $q\coloneqq (\Lambda(p),\Gamma(p)\cup\{\act^n(t_1,t_2)\})$ and notice that $d(p,q)<\omega$.
We have $\A\in\Mod(q)$, which means $q\in P$.
\item This case is similar to the second one.
\item 
Let $\A\in\Mod(p)$.
Since $p\models\Exists{X}\phi$, we have $\A\models \Exists{X}\phi$. 
Let $\B$ be an expansion of $\A$ to $\Delta(p)[X]$ such that $\B\models\phi$.
Since $X$ is finite, there exists a sort preserving injection $\vartheta:X\to C\setminus C_{\Lambda(p)}$.
Let $C_1=\vartheta(X)$ and let $\vartheta':X\to C_1$ be the co-restriction of $\vartheta:X\to C\setminus C_{\Lambda(p)}$ to $C_1$. 
There exists a bijection $\chi:\Delta(p)[X]\to \Delta(p)[C_1]$ which extends $\vartheta':X\to C_1$ and preserves $\Delta(p)$.
Let $\D\coloneqq\Mod(\chi)^{-1}(\B)$. 
Since $\B\models\phi$, by satisfaction condition, 
$\D\models \chi(\phi)$.
Since $\D\red_{\Delta(p)}=\A$ and $\A\models \Gamma(p)$, 
by satisfaction condition,
$\D\models\Gamma(p)$.
\begin{center}
\begin{tikzcd}
\B \ar[r,dotted,no head] & \Delta(p)[X] \ar[rr,"\chi"] & & \Delta(p)[C_1]=\Omega(\ell) & \ar[l,dotted,no head] \D \\
& \A \ar[r,dotted,no head] & \ar[ul,hook] \Delta(p) \ar[ur,hook] &
\end{tikzcd}
\end{center}
By our assumptions there exists $\ell>\Lambda(p)$ such that 
\begin{enumerate*}[label=(\alph*)]
\item $\Omega(\ell)=\Delta(p)[C_1]$, and
\item $\D\in\M(\ell)$. 
\end{enumerate*}
Let $q\coloneqq(\ell,\Gamma(p)\cup\{\chi(\phi)\})$ and notice that $d(p,q)<\omega$.
Since $\D\models\Gamma(p)$ and $\D\models \chi(\phi)$, 
we have $\D\in \Mod(q)$, which means that $q$ is well-defined.
\end{enumerate}
\end{proof}
Lemma~\ref{lemma:up} lays the foundation for proving the following important theorem, which establishes the completeness of the forcing relation with respect to the satisfaction relation.
%%% Semantic Forcing Theorem %%%
\begin{proof}[Proof of Theorem~\ref{th:sfp} (Semantic Forcing Theorem)]
We proceed by induction on the structure of $\phi$.
\begin{description}
\item[$\varphi\in\Sen_0(\Delta(p))$]

To show $\ref{fst1}\Rightarrow\ref{fst2}$, we assume $p\models\varphi$.
\begin{proofsteps}{17em}
let $q\geq p$ be arbitrarily fixed & \\
$\varphi\in f(p)$ & by Lemma~\ref{lemma:sfp}(\ref{lemma:sfp-1}), since $p\models\varphi$ and $\varphi\in\Sen_0(\Delta(p))$\\
$p\Vdash \varphi$ & by the definition of $\Vdash$\\
$q\Vdash \varphi$ & since $p\leq q$\\
$p\Vdash^{w}\varphi$ & since $d(q,q)=0$ and $q\Vdash\varphi$
\end{proofsteps}

To show $\ref{fst2}\Rightarrow\ref{fst1}$, we assume $p\Vdash^w\varphi$. Suppose a contradiction that $p\not\models\varphi$.
\begin{proofsteps}{21em}
$\A\models \neg\varphi$ for some $\A\in\Mod(p)$ & since $p\not\models\varphi$ \\
$\A\in \M(\Lambda(p))\cap (\Gamma(p)\cup\{\neg\varphi\})^\bullet$ & since $\A\in\Mod(p)$ and $\A\models \neg\varphi$ \\
$q\coloneqq (\Lambda(p),\Gamma(p)\cup\{\neg\varphi\})$ belongs to $P$  & since $\A\in\Mod(q)$ \\
$r\not \models \varphi$ for all $r\geq q$ & since $\neg\varphi\in \Gamma(q)$\\
$\varphi \not\in f(r)$ for all $r\geq q$ & by Lemma~\ref{lemma:sfp}(\ref{lemma:sfp-1}) \\
$r\not \Vdash \varphi$ for all $r\geq q$ &  by the definition of $\Vdash$\\
$p\not\Vdash^w\varphi$ & since $q\geq p$ and $r\not\Vdash\varphi$ for all $r\geq q$\\
contradiction & since $p\not\Vdash^w\varphi$ and $p\Vdash^w\varphi$
\end{proofsteps}
\item[$(\act_1\comp\act_2)(t_1,t_2)$] 
To show $\ref{fst1}\Rightarrow\ref{fst2}$, we assume that $p\models (\act_1\comp\act_2)(t_1,t_2)$.
\begin{proofsteps}{20em}
let $q\geq p$ be arbitrarily fixed & \\
$q\models (\act_1\comp\act_2)(t_1,t_2)$ & by Lemma~\ref{lemma:sfp}(\ref{lemma:sfp-2})\\
$r\models \act_1(t_1,c)$ and $r\models \act_2(c,t_2)$ for some $c\in C$ and some 
$r \geq q$ such that $d(q,r)<\omega$ & 
by  Lemma~\ref{lemma:up}(\ref{lemma:up1})\\
$r\Vdash^w \act_1(t_1,c)$ and $r\Vdash^w \act_2(c,t_2)$ & 
by induction hypothesis \\
$u\Vdash \act_1(t_1,c)$ and $u\Vdash \act_2(c,t_2)$ for some $u\geq r$ such that $d(r,u)<\omega$ & 
by the definition of $\Vdash^w$ \\
$u\Vdash(\act_1\comp\act_2)(t_1,t_2) $ and $ d(q,u)<\omega$ & by the definition $\Vdash$ and $d$ \\
$p\Vdash^w(\act_1\comp\act_2)(t_1,t_2)$ & 
since for each $q\geq p$ there exists $u\geq q$ s.t. $d(q,u)<\omega$ and $u\Vdash (\act_1\comp\act_2)(t_1,t_2)$
\end{proofsteps}

To show $\ref{fst2}\Rightarrow\ref{fst1}$, we assume that $p\Vdash^w (\act_1\comp\act_2)(t_1,t_2)$.
\begin{proofsteps}{20em}
let $q\geq p$ be an arbitrary condition & \\
there exists $r\geq q$ such that $r\Vdash(\act_1\comp\act_2)(t_1,t_2)$ & 
since $p\Vdash^w (\act_1\comp\act_2)(t_1,t_2)$\\
$r\Vdash \act_1(t_1,t)$ and $r\Vdash\act_2(t,t_2)$ \newline
for some $t\in T_{\Delta(r)}$ &
by the definition of $\Vdash$\\
$p\Vdash^w\act_1(t_1,t)$ and $p\Vdash^w\act_2(t,t_2)$ &
since for all $q\geq p$ there exist $r\geq q$ and $t\in T_{\Delta(r)}$ s.t. $r\Vdash \act_1(t_1,t)$ \& $r\Vdash\act_2(t,t_2)$\\
$p\models \act_1(t_1,t)$ and $p\models\act_2(t,t_2)$ & 
by induction hypothesis\\
$p\models (\act_1\comp\act_2)(t_1,t_2)$ & 
by semantics
\end{proofsteps}
\item[$(\act_1\cup\act_2)(t_1,t_2)$]
To show $\ref{fst1}\Rightarrow\ref{fst2}$, we assume that $p\models (\act_1\cup\act_2)(t_1,t_2)$.
\begin{proofsteps}{20em}
let $q\geq p$ be arbitrarily fixed & \\
$q\models (\act_1\cup\act_2)(t_1,t_2)$ & by Lemma~\ref{lemma:sfp}(\ref{lemma:sfp-2})\\
$r\models \act_1(t_1,t_2)$ or $r\models \act_2(t_1,t_2)$ for some
$r > q$ such that $d(q,r)<\omega$  & 
by  Lemma~\ref{lemma:up}(\ref{lemma:up2})\\
$r\Vdash^w \act_1(t_1,t_2)$ or $r\Vdash^w \act_2(t_1,t_2)$ & 
by induction hypothesis\\
$u\Vdash \act_1(t_1,t_2)$ or $u\Vdash \act_2(t_1,t_2)$ for some $u\geq r$ such that $ d(r,u)<\omega$ & 
by the definition of $\Vdash^w$, since $r\geq r$\\
$u\Vdash(\act_1\cup\act_2)(t_1,t_2)$ and $d(q,u)<\omega$ & by the definition $\Vdash$ and $d$ \\
$p\Vdash^w(\act_1\cup\act_2)(t_1,t_2)$ & 
since for each $q\geq p$ there exists $u\geq q$ s.t. $d(q,u)<\omega$ and $u\Vdash (\act_1\cup\act_2)(t_1,t_2)$
\end{proofsteps}
To show $\ref{fst2}\Rightarrow\ref{fst1}$, 
we assume that $p\Vdash^w (\act_1\cup\act_2)(t_1,t_2)$.
\begin{proofsteps}{17em}
let $q\geq p$ be an arbitrary condition & \\
$r\Vdash(\act_1\cup\act_2)(t_1,t_2)$ for some $r\geq q$ & 
since $p\Vdash^w (\act_1\cup\act_2)(t_1,t_2)$\\
$r\Vdash \act_1(t_1,t_2)$ or $r\Vdash\act_2(t_1,t_2)$ &
by the definition of $\Vdash$\\
$p\Vdash^w \act_1(t_1,t_2)$ or $p\Vdash^w\act_2(t_1,t_2)$ &
since for all $q\geq p$ there exists $r\geq q$ such that $r\Vdash \act_1(t_1,t_2)$ or $r\Vdash\act_2(t_1,t_2)$\\
$p\models \act_1(t_1,t_1)$ or $p\models\act_2(t_1,t_2)$ & 
by induction hypothesis\\
$p\models (\act_1\cup\act_2)(t_1,t_2)$ & 
by semantics
\end{proofsteps}
\item[$\act^*(t_1,t_2)$] 
To show $\ref{fst1}\Rightarrow\ref{fst2}$, we assume that $p\models \act^*(t_1,t_2)$.
\begin{proofsteps}{22em}
let $q\geq p$ be arbitrarily fixed & \\
$q\models\act^*(t_1,t_2)$ & by Lemma~\ref{lemma:sfp}(\ref{lemma:sfp-2})\\
$r\models \act^n(t_1,t_2)$ for some $n<\omega$ and some $r\geq q$ such that $d(q,r)<\omega$  & 
by  Lemma~\ref{lemma:up}(\ref{lemma:up3})\\
$r\Vdash^w \act^n(t_1,t_2)$ & 
by induction hypothesis \\
$u\Vdash \act^n(t_1,t_2)$ for some $u\geq r$ s.t. $d(r,u)<\omega$ & 
by the definition of $\Vdash^w$\\
$u\Vdash\act^*(t_1,t_2)$ and $d(q,u)<\omega$ & by the definition $\Vdash$ and $d$ \\
$p\Vdash^w\act^*(t_1,t_2)$ & 
as for each $q\geq p$ there exists $u\geq q$ s.t. $d(q,u)<\omega$ and $u\Vdash \act^*(t_1,t_2)$
\end{proofsteps}
To show $\ref{fst2}\Rightarrow\ref{fst1}$, we assume that $p\Vdash^w \act^*(t_1,t_2)$.
\begin{proofsteps}{15em}
let $q\geq p$ be an arbitrary condition & \\
$r\Vdash\act^*(t_1,t_2)$ for some $r\geq q$ & 
since $p\Vdash^w \act^*(t_1,t_2)$\\
$r\Vdash \act^n(t_1,t_2)$ for some $n<\omega$ &
since $r\Vdash \act^*(t_1,t_2)$\\
$p\Vdash^w \act^n(t_1,t_2)$ & 
since for all $q\geq p$ there exists $r\geq q$ s.t. $r\Vdash \act^n(t_1,t_2)$\\
$p\models \act^n(t_1,t_2)$ & 
by induction hypothesis\\
$p\models \act^*(t_1,t_2)$ & 
by semantics
\end{proofsteps}
\item [$\neg\phi$]
For $\ref{fst1}\Rightarrow\ref{fst2}$, 
assume that $p\models\neg\phi$.
For each $q\geq p$,
we show that $q\Vdash\neg\phi$.
\begin{itemize}[topsep=3pt]
\item[] 
Let $q\geq p$.
Suppose towards a contradiction that $q\not\Vdash\neg\phi$.
By the definition of forcing relation, there exists $r\geq q$ such that $r\Vdash\phi$.
It follows that $r\Vdash^w \phi$. 
By induction hypothesis, $r\vDash \phi$.
Since $p\models\neg\phi$ and $r\geq q\geq p$, 
we get $r\vDash \neg \phi$, which is a contradiction.
\end{itemize}
Hence, $p\Vdash^w\neg \phi$.

To show $\ref{fst2}\Rightarrow\ref{fst1}$, assume that $p\Vdash^w\neg\phi$.
Suppose towards a contradiction that $p\not\models\neg\phi$.
\begin{proofsteps}{18em}
there exists $\A\in\Mod(p)$ such that $\A\models\phi$ &
since $p\not\models\neg\phi$\\
let $q=(\Lambda(p),\Gamma(p) \cup \{\phi\}) $ &
$q$ is well-defined, since $\A\in\Mod(q)$\\
$q\models\phi$ &  since $q\models \Gamma(p)\cup\{\phi\}$\\
$q\Vdash^w\phi$ & by induction hypothesis\\
$r\Vdash\phi$ for some $r\geq q$ & 
since $q\geq q$ and $q\Vdash^w\phi$ \\
$u\Vdash\neg\phi$ for some $u\geq r$ & 
since $p\Vdash^w\neg\phi$ and $r\geq p$\\
$u\Vdash\phi$ & since $r\Vdash\phi$ and $u\geq r$\\
contradiction & since $u\Vdash\neg\phi$ and $u\Vdash\phi$
\end{proofsteps}
\item[$\lor\Phi$]
To show $\ref{fst1}\Rightarrow\ref{fst2}$, we assume that $p\models\lor\Phi$.
\begin{proofsteps}{20em}
let $q\geq p$ be an arbitrary condition & \\
$q\models \lor \Phi$ & by Lemma~\ref{lemma:sfp}(\ref{lemma:sfp-2})\\
$r\models \phi$ for some $\phi\in\Phi$ and some $r\geq q$ such that $d(q,r)<\omega$ & 
by  Lemma~\ref{lemma:up}(\ref{lemma:up4})\\
$r\Vdash^w\phi$ & by induction hypothesis\\
$u\Vdash \phi$ for some $u\geq r$ such that $d(r,u)<\omega$ & 
by the definition of $\Vdash^w$\\
$u\Vdash\lor\Phi$ and $d(q,u)<\omega$ & by the definition of $\Vdash$ and $d$ \\
$p\Vdash^w\lor\Phi$ & 
since for all $q\geq p$ there exists $u\geq q$ such that $d(q,u)<\omega$ and $u\Vdash\lor\Phi$
\end{proofsteps}
To show $\ref{fst2}\Rightarrow\ref{fst1}$, we assume $p\Vdash^w\lor \Phi$.
\begin{proofsteps}{17em}
let $q\geq p$ be an arbitrary condition & \\
$r\Vdash\lor\Phi$ for some $r\geq q$ & 
since $p\Vdash^w\lor \Phi$ \\
$r\Vdash\phi$ for some $\phi\in\Phi$ & 
by the definition of $\Vdash$ \\
$p\Vdash^w\phi$ & since for all $q\geq p$ there exists $r\geq q$ s.t. $r\Vdash\phi$\\
$p\models\phi$ & by induction hypothesis \\
$p\models\lor\Phi$ & by semantics 
\end{proofsteps}
\item[$\Exists{X}\phi$]
To show $\ref{fst1}\Rightarrow\ref{fst2}$, we assume that $p\models\Exists{X}\phi$.
\begin{proofsteps}{22em}
let $q\geq p$ be an arbitrary condition & \\
$q\models \Exists{X}\phi$ & by Lemma~\ref{lemma:sfp}(\ref{lemma:sfp-2})\\
$r\models \chi(\phi)$ 
for some bijection $\chi:\Delta(q)[X]\to\Delta(r)$ and 
some $r\geq q$ such that $d(q,r)<\omega$ & 
by  Lemma~\ref{lemma:up}(\ref{lemma:up5})\\
$r\Vdash^w \chi(\phi)$ & by induction hypothesis\\
$u\Vdash\chi(\phi)$ for some $u\geq r$ such that $d(r,u)<\omega$ & 
by the definition of $\Vdash^w$\\
$u\Vdash\Exists{X}\phi$ and $d(q,u)<\omega$ & by the definition of $\Vdash$ and $d$ \\
$p\Vdash^w\Exists{X}\phi$ & since for all $q\geq p$ there exists $u\geq q$ such that $d(q,u)<\omega$ and $u\Vdash\Exists{X}\phi$
\end{proofsteps}

To show $\ref{fst2}\Rightarrow\ref{fst1}$, we assume that $p\Vdash^w \Exists{X}\phi$.
\begin{proofsteps}{20em}
let $q\geq p$ be an arbitrary condition & \\
there exists $r\geq q$ such that $r\Vdash\Exists{X}\phi$ & 
since $p\Vdash^w \Exists{X}\phi$\\
$r\Vdash \theta(\phi)$ 
for some $\theta:X\to \emptyset$ &
by the definition of $\Vdash$\\
$p\Vdash^w\theta(\phi)$ &
since for all $q\geq p$ there exist $r\geq q$ such that $r\Vdash \theta(\phi)$\\
$p\models \theta(\phi)$ & 
by induction hypothesis\\
$p\models \Exists{X}\phi$ & 
by semantics
\end{proofsteps}
\end{description}
\end{proof}
%%% %%% %%% %%% %%% %%% %%% %%% %%%
\section{Proofs of the results presented in Section~\ref{sec:DLS}}
\begin{proof}[Proof of Theorem~\ref{th:DLS} (Downward L\"owenheim-Skolem Theorem)]
Let  $\mathbb{P}(\Omega,\M)$ be a semantic forcing property such that the functors $\Omega$ and $\M$ are defined as in Example~\ref{ex:dls}.
Let $p=((\Sigma,\A),\emptyset)$ be the least condition of $\mathbb{P}(\Omega,\M)$.
By Lemma~\ref{lemma:DLS}, $p$ belongs to a generic set $G$.
Let $\mu:\Delta\Rightarrow \Delta_G$ be a co-limit of $\Delta:(G,\leq)\to \Sig$ such that $\mu_q:\Delta(q)\hookrightarrow\Delta_G$ is an inclusion for all conditions $q\in G$.
By Theorem~\ref{th:gm}, there exists a reachable generic $\Delta_G$-model $\A_G$ for $G$.
For each $p=(\Lambda(p),\Gamma(p))\in G$, we let $\A_p$ denote the unique model from $\M(\Lambda(p))$.
Note that $\A_p$ can be regarded as a pair $(\A,g_p:C_{\Lambda(p)}\to \A)$, 
consisting of the model $\A$ and an interpretation of the constants in $C_{\Lambda(p)}$ into the elements of $\A$ given by the function $g_p$.
For all $(p\leq q)\in (G,\leq)$, we have $\A_q\red_{\Delta(p)}=\A_p$, which means $g_q|_{C_{\Lambda(p)}}=g_p$.
It follows that  $g \coloneqq \bigcup_{p\in G}g_p$ and $\B\coloneqq (\A,g)$ are well-defined.
We show that $\A_G$ is an elementary submodel of $\B$.
Since $\A_G$ is reachable, it suffices to show that $\A_G\models\phi$ implies $\B\models\phi$, for all sentences $\phi\in\Sen(\Delta_G)$.
\begin{proofsteps}{15em}
assume that $\A_G\models\phi$ & where $\phi\in\Sen(\Delta_G)$\\
$q\Vdash \phi$ for some $q\in G$ & $\A_G$ is a generic model for $G$\\
$q\Vdash^w\phi$ & since $q\Vdash \phi$\\
$q\models \phi$ & Theorem~\ref{th:sfp}\\
$\A_q\models \phi$ & since $\Mod(q)=\{\A_q\}$ \\
$\B\models\phi$ & by satisfaction condition, since $\B\red_{\Delta(q)}=\A_q$ and $\A_q\models \phi$ 
\end{proofsteps}
\end{proof}
%%% %%% %%% %%% %%% %%% %%% %%% %%%
\section{Proof of the results presented in Section~\ref{sec:OTT}}
%%% %%% %%% %%% %%% %%% %%% %%% %%%
\begin{proof}[Proof of Theorem~\ref{th:OTT} (Omitting Types Theorem)]
Let  $\mathbb{P}(\Omega,\M)=(P,\leq,\Delta,f)$ be a semantic forcing property such that the functors $\Omega$ and $\M$ are defined as in Example~\ref{ex:ott}.
Let $\Phi$ be a satisfiable set of $\Sigma$-sentences.
Let $\{T_i\subseteq \Sen(\Sigma[X_i])\}_{i<\alpha}$ be a family of types for $\Sigma$ such that $\Phi$ locally omits $T_i$ for all $i<\alpha$. 
We show that there exists a model $\A$ of $\Phi$ which omits  $T_i$ for all $i<\alpha$.
The key point of the proof is to construct a generic set for the least condition $p\coloneqq((\Sigma,\Phi),\emptyset)$ of the semantic forcing property $\mathbb{P}(\Omega,\M)$.
\begin{itemize}[topsep=3pt]
\item Let $triple:\alpha\times\alpha\times\alpha\to\alpha$ be any bijection such that $\beta=triple(i,j,k)$ implies $i \leq \beta$ for all ordinals $i,j,k,\beta<\alpha$.
\item For each condition $q\in P$, let $\gamma(q):\alpha\to \Sen(\Delta(q))$ be an enumeration of the sentences in $\Sen(\Delta(q))$.
\item For each set $D\in\P_\alpha(C)$ and any block of variables $X$ such that $\alpha^{\card(X)}\leq \alpha$,
let $\Theta(X,D)$ denote the set of substitutions from $X$ to $D$.
Furthermore, if $\Theta(X,D)\neq\emptyset$ then we let $\theta(X,D):\alpha\to \Theta(X,D)$ be an enumeration of the substitutions in $\Theta(X,D)$.
\footnote{Since $\alpha^{\card(X_i)}\leq\alpha$ for all $i\in\alpha$, we have $\card(\Theta(X_i,C_{\Lambda(q)}))\leq\alpha$ for all $i<\alpha$ and all $q\in P$.}

\end{itemize}
First, we define an increasing chain of conditions $p_0\leq p_1\leq \ldots \leq p_\beta\leq \ldots$ with the following property:
for all ordinals $\beta<\alpha$ there exists $n<\omega$ such that $d(p,p_\beta)\leq n\cdot\mathrm{card}(\beta)$. 
We proceed by induction on ordinals $\beta<\alpha$.
\begin{description}
\item[($\beta=0$)] Let $p_0\coloneqq p$.
\item[($\beta\Rightarrow\beta+1$)]
Let $(i,j,k)\coloneqq triple^{-1}(\beta)$.
By our assumptions $i\leq\beta$. 
Therefore, the condition $p_i$ is already defined.
\begin{center}
\begin{tikzcd}
T_j  \ar[r,dotted,no head]& \Sen(\Sigma[X_j]) \ar[rr,"{\theta(X_j,C_{\Lambda(p_i)},k)}"] & & 
\Sen(\Delta(p_i)) \ar[r,hook]&  
\Sen(\Delta(p_\beta)) & \ar[l,dotted,no head] \Gamma(p_\beta)\\
& & \ar[ul,hook] \Sen(\Sigma) \ar[ur,hook] & \Phi  \ar[l,dotted,no head]
\end{tikzcd}
\end{center}
\begin{description}
\item[($\Theta(X_j,C_{\Lambda(p_i)})\neq\emptyset$)]
Since $\Phi$ locally omits $T_j$, 
$\Phi\cup \Gamma(p_\beta)\not\models_{\Delta(p_\beta)}\theta(X_j,C_{\Lambda(p_\beta)},k)(T_j)$.

It follows that $\Phi\cup\Gamma(p_\beta)\cup\{\neg\psi\} \not\models_{\Delta(p_\beta)}\bot$  for some $\psi\in\theta(X_j,C_{\Lambda(p_i)},k)(T_j)$.\

Let $q\coloneqq (\Lambda(p_\beta),\Gamma(p_\beta)\cup\{\neg\psi\})$.

Since $\Mod(q)=\Phi^\bullet\cap\Gamma(p_i)^\bullet\cap \{\neg \psi\}^\bullet\neq \emptyset$, we have $q\in P$.
\item[($\Theta(X_j,C_{\Lambda(p_i)})=\emptyset$)] 
Let $q\coloneqq p_\beta$.
\end{description}
We define $p_{\beta+1}$ as follows:
\begin{description}
\item [($r\not\Vdash \gamma(p_i,j)$ for every $r\geq q$)] 
Let $p_{\beta+1}\coloneqq q$. 
In this case, $p_{\beta+1}\Vdash \neg\gamma(p_i,j)$.
\item[($r\Vdash\gamma(p_i,j)$ for some $r\geq q$)] 
Since $r\Vdash^w\gamma(p_i,j)$, by Theorem \ref{th:sfp}, $r\models \gamma(p_i,j)$.
It follows that $\Phi^\bullet\cap \Gamma(r)^\bullet \models_{\Delta(r)} \gamma(p_i,j)$ and
$\Phi^\bullet\cap \Gamma(r)^\bullet \cap \{\gamma(p_i,j)\}^\bullet\neq \emptyset$.
Since $\Gamma(q)\subseteq \Gamma(r)$, we have $\Phi^\bullet\cap \Gamma(q)^\bullet \cap \{\gamma(p_i,j)\}^\bullet\neq \emptyset$.
Therefore, $u\coloneqq (\Lambda(q),\Gamma(q)\cup \{\gamma(p_i,j)\})$ is well-defined. 
Since $u\models\gamma(p_i,j)$, by Theorem~\ref{th:sfp}, we obtain $u\Vdash^w \gamma(p_i,j)$.
There exists $p_{\beta+1}\geq u$ such that $d(u,p_{\beta+1})<\omega$ and $p_{\beta+1}\Vdash \gamma(p_i,j)$.
Since $d(p_\beta,u)<\omega$ and $d(u,p_{\beta+1})<\omega$, we get $d(p_\beta,p_{\beta+1})<\omega$.
By induction hypothesis, $d(p,p_\beta)<n_\beta \cdot \card(\beta)$ for some $n_\beta\in\omega$, which implies $d(p,p_{\beta+1})<n_{\beta+1} \cdot \card(\beta+1)$ for some $n_{\beta+1}\in \omega$.
\end{description}
\item[($\beta<\alpha$ is a limit ordinal)]
Since $(\Delta(p_i),\Gamma(p_i)\cup \Phi)$ is satisfiable for all $i<\beta$,
by compactness, 
$(\bigcup_{i<\beta} \Delta(p_i), (\bigcup_{i<\beta} \Gamma(p_i)) \cup \Phi)$ is satisfiable as well.
Since $\card(C_{\Lambda(p_i)})\leq\card(\beta)$ for all $i<\beta$,
we get $\card(C_{\Lambda(p_\beta)})\leq\card(\beta)<\alpha$.
Similarly, $\card(\Gamma(p_\beta))\leq\card(\beta)<\alpha$.

Hence, $p_\beta\coloneqq\bigcup_{i<\beta}p_i\in P$ is well-defined.
\end{description}
Secondly, we define a generic set $G\coloneqq\{ q\in P\mid q\leq p_\beta \text{ for some }\beta<\alpha\}$.
Let $C_G\coloneqq \bigcup_{\beta<\alpha}C_{\Lambda(p_\beta)}$ and
$\Delta_G\coloneqq\Sigma[C_G]$.
We show that $G$ is a generic set:
\begin{proofsteps}{17em}
let $\phi\in \Sen(\Delta_G)$ & \\
$\phi\in \Sen(\Delta(p_i))$ for some $i<\alpha$ &
since $\Delta_G=\Sigma[C_G]$ and $C_G = \bigcup_{\beta<\alpha}C_{\Lambda(p_\beta)}$ \\
$\phi=\gamma(p_i,j)$ for some $j<\alpha$ & 
since $\gamma(p_i)$ is an enumeration of $\Sen(\Delta(p_i))$\\
let $\beta\coloneqq triple(i,j,k)$ for some $k<\alpha$ & \\
$p_{\beta+1}\Vdash \gamma(p_i,j)$ or $p_{\beta+1}\Vdash \neg \gamma(p_i,j)$ & 
by the construction of  $p_0\leq p_1\leq\ldots$\\
$G\Vdash \phi$ or $G\Vdash \neg\phi$ &
since $\phi=\gamma(p_i,j)$ and $p_{\beta+1}\in G$
\end{proofsteps}
By Theorem~\ref{th:gm}, there exists a reachable generic $\Delta_G$-model $\A_G$ for $G$.
We show $\A_G\red_\Sigma\models\Phi$:
\begin{proofsteps}{17em}
$p_0\Vdash^w \phi$ for all $\phi\in \Phi$ & by Theorem~\ref{th:sfp}, since $p_0\models\phi$ for all $\phi\in \Phi$\\
$G\Vdash\phi$ for all $\phi\in\Phi$ & since $G$ is generic and $p_0\Vdash^w \phi$ for all $\phi\in \Phi$\\
$\A_G\models \phi$ for all $\phi\in\Phi$ & since $\A_G$ is a generic model for $G$\\
$\A_G\red_\Sigma\models\Phi$ & by satisfaction condition, since $\A_G\models\Phi$
\end{proofsteps}
Let $j<\alpha$ be any ordinal.
We show that $\A_G\red_\Sigma$ omits $T_j$.
\begin{proofsteps}{21em}
let $\B$ be an expansion of $\A_G\red_\Sigma$ to $\Sigma[X_j]$ & \\
$\A_G\red_\vartheta=\B$ for some subst. $\vartheta:X_j\to C_G$ &
since $\A_G$ is reachable over $\Delta_G$\\
$\vartheta(X_j)\subseteq C_{\Lambda(p_i)}$ for some $i<\alpha$ &
since $\card(X_j)< cf(\alpha)$\\
let \rlap{$\vartheta': X_j \to C_{\Lambda(p_i)}$ be the co-restriction of $\vartheta:X_j\to C_G$ to $C_{\Lambda(p_i)}$} & \\
$\vartheta'=\theta(X_j,C_{\Lambda(p_i)},k)$ for some $k<\alpha$ &
since $\theta(X_j,C_{\Lambda(p_i)})$ is an enumeration of all substitutions from $X_j$ to $C_{\Lambda(p_i)}$\\
let $\beta=triple(i,j,k)$ & \\
$\neg\psi\in \Gamma(p_{\beta+1})$ for some $\psi\in\theta(X_j,C_{\Lambda(p_i)},k)(T_j)$ & 
by the construction of $p_0\leq p_1\leq\ldots$\\
$p_{\beta+1}\models \neg\psi$ & by semantics\\
$p_{\beta+1}\Vdash^w\neg\psi$ & by Theorem~\ref{th:sfp}\\
$G\Vdash \neg\psi$ & since $G$ is generic and $p_{\beta+1}\in G$\\
$\A_G\models \neg\psi$ & since $\A_G$ is generic for $G$\\
$\A_G\not\models\theta(X_j,C_{\Lambda(p_i)},k)(T_j)$ & since $\psi\in \theta(X_j,C_{\Lambda(p_i)},k)(T_j)$\\
$\A_G\not\models\vartheta(T_j)$ & since $\vartheta(T_j)=\theta(X_j,C_{\Lambda(p_i)},k)(T_j)$\\
$\B\not\models T_j$ & by satisfaction condition for substitutions\\
$\A_G\red_\Sigma$ omits $T_j$ & since $\B$ is an arbitrary expansion of $\A_G\red_\Sigma$ to $\Sigma[X_j]$ for which we proved that $\B\not\models T_j$
\end{proofsteps}
\end{proof}
%%% %%% %%% %%% %%% %%% %%% %%% %%%
\begin{proof}[Proof of Lemma~\ref{lemma:inf}]
Suppose towards a contradiction that $\Phi\cup \Gamma\models_{\Sigma[y]} T$ for some finite set $\Gamma\subseteq\Sen(\Sigma[y])$ such that $\Phi\cup \Gamma$ is satisfiable over $\Sigma[y]$.
For each $n>0$, let 
\begin{enumerate}
\item $\Sigma_n\coloneqq (\{s_i\mid 0 < i\leq n \},\emptyset,\emptyset)$ be a signature, which is included in $\Sigma$, and
\item  $\Phi_n\coloneqq \{\phi_i\mid 0< i\leq n \}$  a set of $\Sigma_n$-sentences, which is included in $\Phi$.
\end{enumerate}
Since $\Gamma$ is finite, $\Gamma\subseteq \Sigma_k[y]$ for some $k > 0$.
We show that $\Phi_k\cup \Gamma\models_{\Sigma_k[y]} T$:
\begin{proofsteps}{20em}
let $\A$ be a model of $(\Sigma_k[y],\Phi_k\cup \Gamma)$ & \\
let $\B$ be the expansion of $\A$ to $\Sigma[y]$ \rlap{ which interprets $s_n$ as $\emptyset$ for all $n > k$} & \\
$\B\models \Phi_k\cup\Gamma$ & by satisfaction condition, since $\A\models \Phi_k\cup\Gamma$\\
$\B\models\phi_n$ for all $n>k$ & since $\B_{s_n}=\emptyset$ for all $n>k$\\
$\B\models \Phi$ & since $\B\models \Phi_k$ and $\B\models\phi_n$ for all $n>k$\\
$\B\models T$ & since $\Phi\cup \Gamma\models_{\Sigma[y]} T$
\end{proofsteps}
\noindent Since $\Phi_k\cup \Gamma$ is finite and $\Phi_k\cup \Gamma\models_{\Sigma_k[y]} T$, 
it follows that infinity can be modeled by a finite set of sentences, which is a contradiction with compactness.
Hence, $\Phi$ locally omits $T$.
\end{proof}
Let $\Sigma$, $\Phi$ and $T$ be as defined in Lemma~\ref{lemma:inf}.
Recall that $C=\{C_{s_n}\}_{n<\omega}$ is the set of all constants, and $\card(C_{s_n})=\omega$ for all natural numbers $n$. 
Notice that $(\Sigma,\Phi)$ is satisfiable and by Lemma~\ref{lemma:inf}, $\Phi$ locally omits $T$. 
By Theorem~\ref{th:OTT}, there exists a model $\A$ for $\Phi$ which omits $T$.
Since any model of $(\Sigma[C],\Phi)$ realizes $T$, the model $\A$ is constructed from a proper subset of $C$  in contrast to classical first-order logic, where results are developed over the entire $\Sigma[C]$.
%%% %%% %%% %%% %%% %%% %%% %%% %%%
\section{Proofs of the results presented in Section~\ref{sec:complete}}

\begin{lemma}\label{lemma:list}
Let $\Sigma$ be the signature defined in Example~\ref{ex:list}.
Let $\Phi$ be a set of $\Sigma$-sentences which defines $add:List~List\to List$ by induction on the second argument:
\begin{enumerate*}[label=(\alph*)]
\item $\Forall{x}add(x,empty)=x$ and
\item $\Forall{x,y,e} add(x,y;e)=add(x,y);e$
\end{enumerate*} 
where $x$ and $y$ are variables of sort $List$ while $e$ is a variable of sort $Elt$.
\begin{enumerate}
\item $\Phi\models^c\Forall{x,y,z}add(add(x,y),z)=add(x,add(y,z))$ and
\item $\Phi\not\models\Forall{x,y,z}add(add(x,y),z)=add(x,add(y,z))$, 
\end{enumerate}
where $x$, $y$ and $z$ are variables of sort $List$.
\end{lemma}
\begin{proof}[Proof of Lemma~\ref{lemma:list}]
First, 
notice that the following proof rule is sound in $\frag^c$.
\begin{longtable}{l l l}
$(CB)~\Frac{\Phi\vdash^c \Forall{var(t)}\psi(t) \text{ for all }  t\in T_{\Sigma^c}(Y)}{\Phi \vdash^c \Forall{x}\psi(x)}$\\
\end{longtable}
\noindent where 
\begin{enumerate*}[label=(\alph*)]
\item $x$ is a variable of constrained sort,
\item $Y$ is a $S^l$-sorted block of variables such that $\card(Y_s)=\omega$ for all loose sorts $s\in S^l$, and
\item $var(t)$ is the set of all variables occuring in $t$.
\end{enumerate*}
$(CB)$ asserts that to prove the property $\psi$ for an arbitrary element $x$, it suffices to establish $\psi$ for all constructor terms $t$ in $T_{\Sigma^c}(Y)$.

Then prove that $\Phi\models^c\Forall{x,y,z}add(add(x,y),z)=add(x,add(y,z))$, where $x$, $y$ and $z$ are variables of sort $List$.
Let $Y$ be a countably infinite block of variables of sort $Elt$.
We show that 
$\Phi\models_{\Sigma[\overline{x},\overline{y},\overline{z}]} add(add(\overline{x},\overline{y}),\overline{z})=add(\overline{x},add(\overline{y},\overline{z}))$,
where 
$\overline{x}$, $\overline{y}$ and $\overline{z}$ are lists of variables from $Y$, that is,
\begin{enumerate*}[label=(\alph*)]
\item $\overline{x}=empty;x_1;\dots;x_n$, where $x_i\in Y$ for all $i\in\{1,\dots,n\}$,
\item $\overline{y}=empty;y_1;\dots;y_m$, where $y_i\in Y$ for all $i\in\{1,\dots,m\}$ and
\item $\overline{z}=empty;z_1;\dots;z_k$, where $z_i\in Y$ for all $i\in\{1,\dots,k\}$.
\end{enumerate*}
We proceed by induction on the structure of $\overline{z}$:
\begin{enumerate}[label=(\alph*)]
\item Since $\Phi\models_{\Sigma[\overline{x},\overline{y}]} add(add(\overline{x},\overline{y}),empty)=add(\overline{x},\overline{y})$ and 

$\Phi\models_{\Sigma[\overline{x},\overline{y}]} add(\overline{x},add(\overline{y},empty))=add(\overline{x},\overline{y})$, we get

$\Phi\models_{\Sigma[\overline{x},\overline{y}]}  add(add(\overline{x},\overline{y}),empty)=add(\overline{x},add(\overline{y},empty))$.

\item Since $\Phi\models_{\Sigma[\overline{x},\overline{y},\overline{z},e]} add(add(\overline{x},\overline{y}),\overline{z};e)=add(add(\overline{x},\overline{y}),\overline{z});e$ 
by induction hypothesis, 
$\Phi\models_{\Sigma[\overline{x},\overline{y},\overline{z},e]} add(add(\overline{x},\overline{y}),\overline{z};e)=add(\overline{x},add(\overline{y},\overline{z}));e$.

Since $\Phi\models_{\Sigma[\overline{x},\overline{y},\overline{z},e]} add(\overline{x},add(\overline{y},\overline{z};e))=add(\overline{x},add(\overline{y},\overline{z}));e$,
it follows that 

$\Phi\models_{\Sigma[\overline{x},\overline{y},\overline{z},e]} add(add(\overline{x},\overline{y}),\overline{z};e)=add(\overline{x},add(\overline{y},\overline{z};e))$.  
\end{enumerate}
Since $\Phi\models_{\Sigma[\overline{x},\overline{y},\overline{z}]} add(add(\overline{x},\overline{y}),\overline{z})=add(\overline{x},add(\overline{y},\overline{z}))$,
by semantics, we have 
$\Phi\models_{\Sigma}\Forall{\overline{x},\overline{y},\overline{z}} add(add(\overline{x},\overline{y}),\overline{z})=add(\overline{x},add(\overline{y},\overline{z}))$ 
for all lists $\overline{x}$, $\overline{y}$ and $\overline{z}$ with elements from $Y$.
By soundness of $(CB)$, we obtain $\Phi\models^c\Forall{x,y,z}add(add(x,y),z)=add(x,add(y,z))$, where $x$, $y$ and $z$ are variables of sort $List$.

For the second statement, we construct a $\Sigma$-model $\B$ such that $\B\models\Phi$ and $\B\not\models\Forall{x,y,z}add(add(x,y),z)=add(x,add(y,z))$.
Let $\Sigma'$ be the signature obtained from $\Sigma^c$ by adding a new constant $nil:\to List$.
The $\Sigma$-model $\B$ is obtained from $T_{\Sigma'}(\omega)\red_{\Sigma^c}$\footnote{This means that $\B_{Elt}=\omega$ and $\B_{List}$ consists of lists of one of the forms $empty$, $(empty;m_1;\dots;m_k)$, $nil$  or $(nil;n_1;\dots;n_l)$, 
where $n_i$ and $m_j$ are natural numbers.} 
by interpreting $add$ as follows:
\begin{enumerate}[label=(\alph*)]
\item $add^\B(\ell,empty)=\ell$,
\item $add^\B(\ell,\ell';n)=add^\A(\ell,\ell');n$,
\item $add^\B(empty,nil)=empty;0$ and
\item $add^\B(\ell;n,nil)=\ell;n$,
\end{enumerate}
where  $\ell,\ell'\in \B_{List}$ and $n\in \B_{Elt}$.
By definition, $\B\models\Phi$.
However, for all $n<\omega$ we have
$add^\B(add^\B(empty;n,empty),nil)= empty;n$ which is different from the list $add^\B(empty;n,add^\B(empty,nil)) = empty;n;0$.
\end{proof}
%%% %%% %%%
\begin{proof}[Proof of Theorem~\ref{th:complete} (Completeness)]
Consider a set of sentences $\Phi$ defined over a signature $(S^f\subseteq S,F^c\subseteq F,L)$.
Let $Y$ be a $S^l$-sorted block of variables such that $\card(Y_s)=\omega$ for all sorts $s\in S^l$. 
Let $\alpha\coloneqq\card(S^f)$ and $\{s_i\}_{i<\alpha}$ is an enumeration of $S^f$.
We say that $\Phi$ is semantically closed under $(CB)$ whenever $\Phi\models_\Sigma \Forall{var(t)} \psi(t)$ for all constructor terms $t\in T_{(S,F^c)}(Y)$ implies $\Phi\models_\Sigma\Forall{x}\psi(x)$.
Similarly, we say that $\Phi$ is semantically closed under $(FN)$ whenever $\Phi\cup\Gamma_{\overline{n}}\models\psi$ for all $\overline{n}\in \omega^\alpha$ implies $\Phi\models\psi$.

First, we assume that $\Phi$ is semantically closed under $(CB)$ and $(FN)$ and satisfiable in $\frag$.
Then we prove that $\Phi$ is satisfiable in $\frag^f$ in three steps:
\begin{enumerate}
\item  $\Phi$ omits  type $T^c\coloneqq\{\Forall{var(t)} x\neq t \mid t\in T_{\Sigma^c}(Y)\}\subseteq \Sen(\Sigma[x])$.
\begin{proofsteps}{17.5em}
let $\gamma$ be a $\Sigma[x]$-sentence such that \rlap{$\Phi\cup\{\gamma(x)\}$ is satisfiable over $\Sigma[x]$} & \\ 
$\Phi\not\models\Forall{x}\neg\gamma(x)$ & since $\Phi\cup\{\gamma(x)\}$ is satisfiable over $\Sigma[x]$\\
$\Phi\not\models \Forall{var(t)}\neg \gamma(t)$ for some $t\in T_{\Sigma^c}(Y)$ & since $\Phi$ is semantically closed under $(CB)$\\
$\Phi\cup\{\gamma(t)\}\not\models_{\Sigma[var(t)]}\bot$  & by semantics\\
$\Phi\cup\{\gamma(x)\}\cup\{\Exists{var(t)} t=x \}\not\models_{\Sigma[x]}\bot$ & as $\A\models\Phi\cup\{\gamma(t)\}$ for some $\A\in|\Mod(\Sigma[var(t)])$\\
$\Phi$ locally omits $T$ & by Lemma~\ref{lemma:otp}
\end{proofsteps}

\item $\Phi$ omits type $T_s^f\coloneqq\{\Exists{x_1,\dots,x_n} \land_{i\neq j} x_i\neq x_j \mid x_1,\dots, x_n \text{ are variables of sort }s\}$, for all sorts $s\in S^f$.
\begin{proofsteps}{17em}
let $\gamma$ be a $\Sigma$-sentence such that \rlap{$\Phi\cup\{\gamma\}$ is satisfiable over $\Sigma$} & \\ 
$\Phi\cup \{\gamma\}\not\models\bot$ & since $\Phi\cup\{\gamma\}$ is satisfiable over $\Sigma$\\
$\Phi\cup \{\gamma\}\cup \Gamma_{\overline{n}}  \not\models \bot$ for some $\overline{n}\in\omega^\alpha$ & since $\Phi$ is semantically closed under $(FN)$\\
$\A\models \Phi\cup \{\gamma\}\cup \Gamma_{\overline{n}}$ for some $\Sigma$-model $\A$  & by semantics\\
$\A\not\models T_s^f$ for all $s\in S^f$ & since $\A\models \Gamma_{\overline{n}}$\\
$\Phi$ locally omits $T_s^f$ for all $s\in S^f$ & by Lemma~\ref{lemma:otp}, since $\A\models \Phi\cup \{\gamma\}$ and $\A\not\models T_s^f$ for all $s\in S^f$
\end{proofsteps}

\item By Theorem~\ref{th:OTT}, there exists a model $\A$ which satisfies $\Phi$ and omits $T^c$ and  $T_s^f$ for all $s\in S^f$.
By the definition of the types $T^c$ and $T_s^f$, the model $\A$ is a constructor-based algebra which interprets each sort $s\in S^f$ as a finite set.
\end{enumerate}

Secondly, we assume that $\Phi\not \vdash^f \bot$ and then we prove that $\Phi\not\models^f\bot$.
We define the following set of sentences $\Psi\coloneqq \{\psi\in\Sen(S^f\subseteq S,F^c\subseteq F,L)\mid \Phi\vdash^f \psi\}$.
We show that $\Phi\not\vdash^f\bot$ iff $\Psi\not\vdash \bot$:
\begin{description}
\item [$\Rightarrow$]
For the forward implication, suppose towards a contradiction that $\Psi$ is not consistent in
$\frag$, that is, $\Psi\vdash\bot$.
Since $\vdash\subseteq \vdash^f$, we get $\Psi\vdash^f\bot$.
By $(Union)$, $\Phi\vdash^f \Psi$.
By $(Transitivity)$, $\Phi\vdash^f\bot$, 
which is a contradiction with the consistency of $\Phi$ in $\frag^f$.

\item [$\Leftarrow$]
For the backward implication, suppose towards a contradiction that $\Phi\vdash^f\bot$.
By the definition of $\Psi$, $\bot\in \Psi$.
By $(Monotonicity)$, $\Psi\vdash \bot$, which is a contradiction with the consistency of $\Psi$ in $\frag$. 
\end{description}
Assume that $\Phi$ is consistent in $\frag^f$. 
By the above property, $\Psi$ is consistent in $\frag$.
By completeness of $\frag$, $\Psi$ is satisfiable in $\frag$.
By completeness and soundness of $\frag$, $\Psi$ is semantically closed under $(CB)$ and $(FN)$.
By the first part of the proof, $\Psi$ is satisfiable in $\frag^f$.
Since $\Phi\subseteq \Psi$, $\Phi$ is satisfiable in $\frag^f$. 
\end{proof}
%%% %%% %%% %%% %%% %%% %%% %%% %%%
\end{document}